\documentclass[10pt,pre,aps,amsmath,amsfonts,amssymb,floatfix,superscriptaddress,notitlepage, twocolumn]{revtex4-1}
\usepackage{epsfig,amssymb,amsfonts,graphicx,color,calc,epstopdf,textcomp}

\let\a=\alpha \let\b=\beta \let\g=\gamma \let\d=\delta
   
\let\l=\lambda    
\let\s=\sigma \let\t=\tau \let\f=\varphi \let\ph=\varphi\let\c=\chi
   \let\G=\Gamma
\let\D=\Delta  \let\L=\Lambda  
   
  \let\th=\theta \let\io=\infty
 
\renewcommand{\vec}[1]{\mathbf{#1}}
\def\vr{{\vec r}}

\def\FF{{\cal F}}

\newcommand{\myref}[1]{(\ref{#1})}

\newcommand{\beq}{\begin{equation}}
\newcommand{\eeq}{\end{equation}}
\newcommand{\ba}{\begin{eqnarray}}
\newcommand{\ea}{\end{eqnarray}}

\newcommand{\av}[1]{\overline{\left\langle {#1} \right\rangle}}

\def\phid{\varphi_{\rm d}}
\def\phij{\varphi_{\rm J}}
\def\phiG{\varphi_{\rm G}}
\def\phiin{\varphi_{\rm 0}}
\def\DEA{\Delta_{\rm EA}}
\def\DAB{\Delta_{AB}}

\newcommand{\br}{\mathbf{r}}

\begin{document}
\title{Numerical detection of the Gardner transition in a mean-field glass former}

\author{Patrick Charbonneau}
\affiliation{Department of Chemistry, Duke University, Durham,
North Carolina 27708, USA}
\affiliation{Department of Physics, Duke University, Durham,
North Carolina 27708, USA}

\author{Yuliang Jin}
\email{yuliang.jin@lpt.ens.fr}
\affiliation{Department of Chemistry, Duke University, Durham,
North Carolina 27708, USA}
\affiliation{Dipartimento di Fisica,
Sapienza Universit\'a di Roma, INFN, Sezione di Roma I, IPFC -- CNR, Piazzale Aldo Moro 2, I-00185 Roma, Italy}
\affiliation{LPT,
\'Ecole Normale Sup\'erieure, UMR 8549 CNRS, 24 Rue Lhomond, 75005 France}

\author{Giorgio Parisi}
\affiliation{Dipartimento di Fisica,
Sapienza Universit\'a di Roma, INFN, Sezione di Roma I, IPFC -- CNR, Piazzale Aldo Moro 2, I-00185 Roma, Italy}

\author{Corrado Rainone}
\affiliation{Dipartimento di Fisica,
Sapienza Universit\'a di Roma, INFN, Sezione di Roma I, IPFC -- CNR, Piazzale Aldo Moro 2, I-00185 Roma, Italy}
\affiliation{LPT,
\'Ecole Normale Sup\'erieure, UMR 8549 CNRS, 24 Rue Lhomond, 75005 France}

\author{Beatriz Seoane}
\email{seoanebb@roma1.infn.it}
\affiliation{Dipartimento di Fisica,
Sapienza Universit\'a di Roma, INFN, Sezione di Roma I, IPFC -- CNR, Piazzale Aldo Moro 2, I-00185 Roma, Italy}
\affiliation{Instituto de Biocomputaci\'on y F\'{\i}sica de Sistemas Complejos (BIFI), 50009 Zaragoza, Spain}

\author{Francesco Zamponi}
\affiliation{LPT,
\'Ecole Normale Sup\'erieure, UMR 8549 CNRS, 24 Rue Lhomond, 75005 France}

\begin{abstract}
Recent theoretical advances predict the existence, deep into the glass phase, of a novel phase transition, the so-called {\it Gardner transition}. 
 This transition is associated with the emergence of a complex free energy landscape composed of many marginally stable % glassy sub-basins. 
{sub-basins within a glass metabasin.}
 In this study, we explore several methods to detect numerically the Gardner transition in a simple structural glass former, the infinite-range Mari-Kurchan model. The transition point is robustly located from three independent approaches: (i) the divergence of the characteristic relaxation time, (ii) the divergence of the caging susceptibility, and (iii) the abnormal tail in the probability distribution function of cage order parameters. 
We show that the numerical results are fully consistent with the theoretical expectation.
 The methods we propose may also be generalized to more realistic numerical models as well as to experimental systems.
\end{abstract}

\maketitle

%\tableofcontents
\section{Introduction}

Upon compressions that are sufficiently rapid to avoid crystallization, a fluid of hard spheres (HS) 
first turns sluggish and then forms a glass~\cite{BB11,Ca09}. 
This glass can then be further compressed until the system jams~\cite{LN98}, which occurs under the application of an infinite confining pressure~\cite{KL07,PZ10}. 
Glass formation is entropic, i.e., particles vibrate and thus cage each other in place, while jamming is mechanical, i.e., no motion is possible and particles are held steady through direct contacts with each other. Over the last decade, this two-transition scenario has been broadly validated, 
both numerically and theoretically~\cite{DTS05,SDST06,SLN06,CBS09,HD09,PZ10,CIPZ11,IBS12,IBS13}. Interestingly, recent 
advances predict that -- at least in the mean-field, infinite-dimensional ($d\to\io$) limit -- there exists 
a third transition, a so-called {\it Gardner transition}, that is intermediate in density and pressure between glass formation and jamming~\cite{KPUZ13,CKPUZ14,Charbonneau2014B,Rainone2014}. 
First discovered in spin-glass models~\cite{Ga85,GKS85,MR03,MR04,Ri13}, the Gardner transition corresponds to 
a single glass {metabasin} %state 
splitting into a complex hierarchy of marginally stable {sub-basins}. % sub-states.  
The transition is thus akin to the spin-glass transition of the Sherrington-Kirkpatrick (SK) model, wherein a critical temperature separates a 
paramagnetic phase, in which a single thermodynamic state exists, from a marginal phase, in which a large number of distinct spin-glass states appear~\cite{MPV87}.
In structural glasses, however, the high-temperature phase corresponds to a given glass {metabasin}
that has been dynamically selected by a quenching protocol; it is this {metabasin} that
then undergoes a spin-glass-like transition~\footnote{
The Gardner transition is akin to that of the SK model in presence of a random magnetic field, because the system is confined in a given glass 
metabasin and the self-induced disorder characteristic of this glass metabasin acts as a self-induced 
``external'' random field. At the mean-field level, this distinction does not make any 
difference, but the presence of a random field is important in finite dimensions~\cite{UB14}.
}.

The discovery of a 
Gardner transition in glasses has already markedly advanced our theoretical understanding of jamming 
by providing analytical predictions for the critical {jamming} exponents~\cite{CKPUZ14,LDW13,Charbonneau2014B,CCPZ15,MW15}.
It further suggests an explanation for the abundance of soft vibrational modes in glasses~\cite{WNW05,MW15},
for the peculiar behavior of the specific heat in quantum glasses~\cite{LDW13,KPUZ13}, 
and various other transport and thermodynamic properties in this regime. 

Before these fascinating problems can be {tackled, however,} %addressed by this approach,
a crucial question is whether the Gardner transition itself, 
whose existence is well established in the $d\rightarrow\io$ limit, exists in finite (low) dimensions. %In fact,
Renormalization group results indicate
that the transition might disappear or dramatically change of nature in low $d$~\cite{UB14}. 
Yet a similar line of inquiry has been pursued for decades in the context of spin glasses~\cite{BY86},  
leading to the conclusion that, whatever the ultimate fate of the phase transition in the thermodynamic limit may be,
the $d\to\io$ scenario provides a very
good description of the system over the relevant experimental length and time scales~\cite{Janus}.
At present, the {most direct} %only 
way to assess the relevance of the Gardner transition for the description of experimental glasses is through %the use of 
numerical simulations.
It is therefore important to {first} identify the observable consequences of this transition {in well-controlled model systems}.

This study primarily aims to develop procedures and to identify observables in order  to reliably detect the Gardner transition. 
To that effect, we consider a simple structural glass former, the infinite-range Mari-Kurchan (MK) model~\cite{MK11,mari:09}.
The model is quite abstract and in some ways far from realistic models of glasses, but
{\it (i)} it is a mean-field model by construction;
{\it (ii)} it shares, in any finite dimension $d$, the same qualitative phase diagram as infinite-dimensional hard spheres, provided one neglects {the effect of hopping on the glassy dynamics}~\cite{Charbonneau2014};
%some hopping effects that affect the dynamics on long time scales
{\it (iii)} it can be studied analytically in great detail, using the methods of~\cite{Rainone2014,Charbonneau2014}, that we further developed
for this work;
{\it (iv)} {and,} most importantly, it can be easily simulated in any finite dimensions $d$, {including $d=3$ as we do here.} %; we focus on $d=3$ in the following.
{Discerning 
%Our aim here is to discern 
the signatures of the Gardner transition in this well-controlled setting, 
where we are certain that the transition exists, shall later on enable us and others to study %, before later considering
more realistic glass models for which the existence of the transition is not a priori guaranteed.}

{The plan of this paper is as follows. In Sect.~\ref{sec:II} we describe the MK model and its glassy behavior, and in Sec.~\ref{sec:numerics} we detail the numerical procedures we use for the study. In Sec.~\ref{sec:GardnerDetection} we discuss several quantities that bear the signature of the transition, as suggested by the analogy with spin glasses.}  
For instance, at both the spin-glass and the Gardner transitions, the ``spin-glass susceptibility'' diverges and the distribution of overlaps (distances) between different replicas becomes non-trivial. {Bringing the various estimates of the transition together in Sec.~\ref{sec:summary} reveals that} %, we find that 
the Gardner transition can be reliably and reproducibly located through numerical methods in the MK model, and {that} the results {are} fully consistent
with theoretical expectations.
{We conclude in Sec.~\ref{sec:conclusion} with a description of} %we also discuss 
other possible measurements to detect and characterize the Gardner transition, which may be more appropriate for numerical simulations of more realistic model glass formers as well as for experiments.

%Let us conclude the introduction by a warning to the reader. 
{Before embarking on this program, let us make a note of warning to the reader.}
Because the aim of work is %mostly 
to identify numerical methods to detect the Gardner
transition, we {have attempted} 
%tried 
to make the numerical part as self-contained as possible. %On the contrary, 
For what concerns the theoretical part, {however, we have chosen to be more succinct and have instead} %in order to keep the paper short enough we 
relied on previous work both on spin glasses and structural glasses, which is {here} only briefly recalled. We {expect} that the reader unfamiliar
with spin-glass theory will {nonetheless} be able to read and understand the numerical part without much difficulty.

\section{Model and basic physical picture}
\label{sec:II}

\begin{figure}[t]
\centerline{\hbox{\includegraphics [width = \columnwidth] {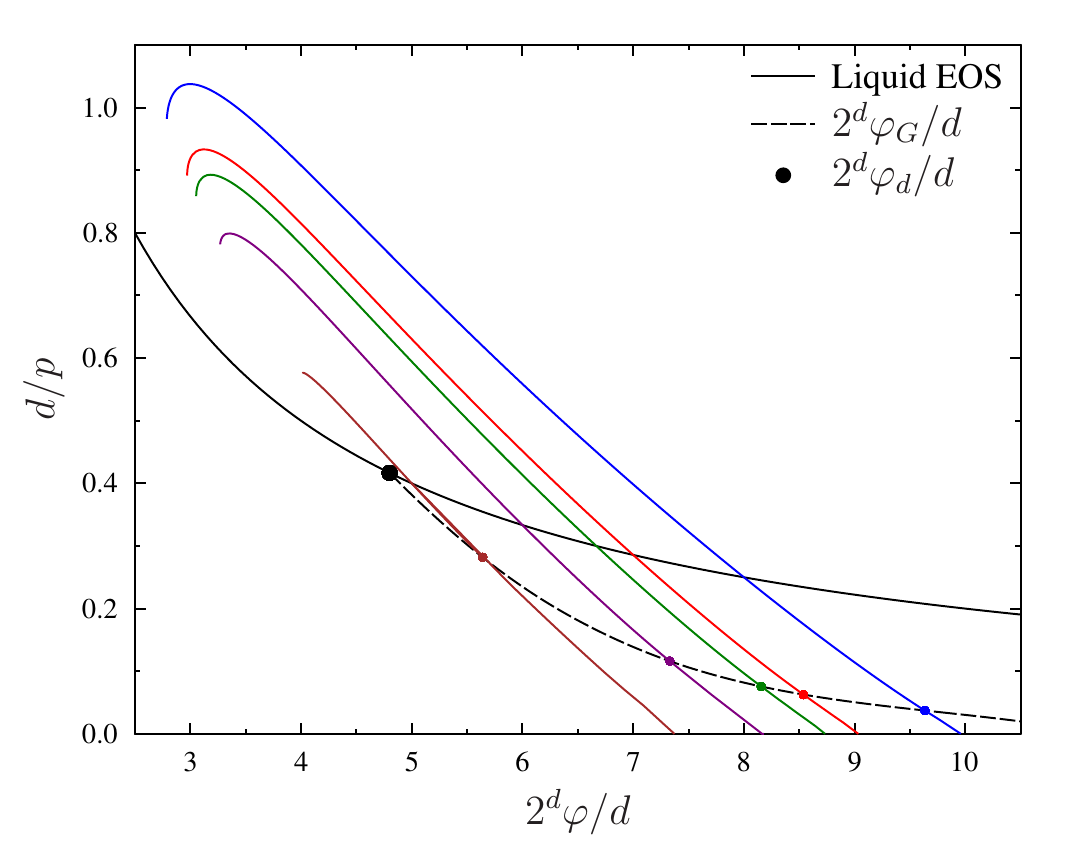}}}
\caption{(Color online).
Phase diagram of the HS and MK models in the limit $d\to\io$. 
Results are partially derived from Ref.~\cite{Rainone2014}, complemented by new results obtained %in this work 
for the Gardner phase{~\cite{Rainone2666}}.
The liquid EOS given by Eq.~\eqref{eq:liquid_eos} (full black line) 
gives $p/d = (2^d \f/d)/2$ in the limit $d\to\io$. The black dot denotes the dynamical glass transition at $2^d\phid/d=4.8$. Glass EOSs from state following (SF) for $2^d \phiin/d = 5, 6, 6.667, 7, 8$ are also reported (thin colored lines, from left to right). The envelope for the Gardner transition for each SF (colored dot) is given by the dashed line. 
}
\label{fig:theory}
\end{figure}

We consider a simple glass-former, the infinite-range Mari-Kurchan model~\cite{MK11,mari:09} -- initially proposed by Kraichnan~\cite{Kr62} -- 
in which  $N$ hard spheres of diameter $\sigma$ interact through a pair potential {that} is a function of distance
shifted by a quenched random vector $\mathbf{\L}_{ij}$. The total interaction energy is thus
 \beq 
 U = \sum_{i<j}^{N} u\left(|\br_{ij}|\right), 
 \eeq
where for particles at positions $\{\br_i\}$ the shifted distance  $\br_{ij}$ is defined as $\br_{ij} =\br_i - \br_j + \mathbf{\L}_{ij}$, 
and $u(r )$ is the HS potential, i.e., $e^{-u(r)} = \th(r -\s)$ with $r = |\br|$. The random {shifts}, which are uniformly distributed over the system volume $V$, 
induce a quenched disorder that suppresses both crystallization and nucleation between metastable glassy states~\cite{MK11,mari:09,Charbonneau2014}. 
The model further enables {\it planting}, which is a 
simple process for generating equilibrated liquid configurations at all densities~\cite{KZ09,Charbonneau2014} (see Sec.~\ref{sec:planting}). 
In all spatial dimensions, the MK model has a mean-field structure \emph{by construction}, 
due to the infinite-range random shifts~\cite{MK11},
and exhibits a jamming transition in the same universality class as standard HS \cite{mari:09}.
The MK model is also fully equivalent to standard HS in the limit $d \to \infty$, where both models can be solved 
exactly using the mean-field methods {described in} Refs.~\cite{PZ10,MK11,KPUZ13,CKPUZ14,Charbonneau2014B,Rainone2014,Rainone2666}.
%In particular, for this study we complemented the {\it state-following} (SF) study of~\cite{Rainone2014} by writing and solving
%the full Replica Symmetry Breaking (fullRSB) equations that are needed to describe the marginal Gardner phase, 
%as will be detailed elsewhere~\cite{Rainone2666}.
In the rest of this section, we briefly describe the phase diagram in the infinite-dimensional limit (Fig.~\ref{fig:theory}), 
we explain how it can be applied to the MK model in $d=3$, and we discuss
some of the finite-dimensional corrections that have thus far been considered~\cite{Charbonneau2014}.

\subsection{Equilibrium states (liquid phase)}

The liquid phase of the MK model ergodically samples equilibrium configurations following the Gibbs distribution and
has a remarkably simple structure. Its pair correlation function is given by
\begin{equation}
\begin{split}
g_2(\br) & \equiv
 \frac{V}{N(N-1)} \overline{\left\langle \sum_{i \neq j} \d( \br_{ij}- \br ) \right\rangle} \\
& = e^{-\beta u(r )} =
\theta (r - \sigma) ,
\end{split}
\label{eq:g2}
\end{equation}
where $\beta = 1/T$ is the inverse temperature, $\langle \cdots \rangle$ denotes thermal averaging, and $\overline{\cdots}$ denotes averaging over
quenched disorder, i.e., over {$\mathbf{\L}_{ij}$}. The second virial coefficient is
\beq
B_2 =- \frac{1}{2V} \iint  f(\br_{12}) d \br_1 d\br_2 = \frac{V_d \sigma^d}{2},
\eeq 
where the Mayer function $f(\br) = e^{-\beta u(r)} - 1$, and  $V_d$ is the volume of a $d$-dimensional ball of unit radius. 

Because no indirect correlations exist, higher-order correlation functions can be factorized in a trivial way and the corresponding virial coefficients are zero. For example, the three-body correlation function %is
\beq
\begin{split}
g_3(\br, \br') & \equiv
 \frac{V^2}{N(N-1)(N-2)} \overline{\left\langle \sum_{i \neq j \neq k} \d( \br_{ij} - \br ) \d( \br_{ik} - \br' ) \right\rangle} \\
 & = g_2(\br) g_2(\br'), 
\end{split}
\eeq
and the third virial coefficient
\beq
\begin{split}
B_3 & = -\frac{1}{3V}\iiint f(\br_{12})f(\br_{13})f(\br_{23}) d\br_1 d\br_2 d\br_3 \\
&= -\frac{1}{3V}\iiint f(\br_{12})  f(\br_{13}) \times \\
& \times f(\br_{13} - \br_{12} + \mathbf{\L}_{12} +  \mathbf{\L}_{23} -   \mathbf{\L}_{13}) d\br_1 d\br_2 d\br_3 \\
& = 0.
\end{split}
\eeq
Note that if $|\br_{12}|, |\br_{13}| < \s$, then
$f(\br_{13} - \br_{12} + \mathbf{\L}_{12} +  \mathbf{\L}_{23} -   \mathbf{\L}_{13})=0$ in the thermodynamic limit because random shifts are uncorrelated and typically of the system size, {and thus} $|\br_{13} - \br_{12} + \mathbf{\L}_{12} +  \mathbf{\L}_{23} -   \mathbf{\L}_{13}| \gg \sigma$. It is straightforward to generalize this argument to show that all higher-order virial coefficients are also zero~\cite{MK11}. 

Because only the second virial coefficient is non-zero, the reduced pressure $p$ equation of state (EOS) for the liquid {is}
\begin{equation}
p \equiv \beta P/\rho = 1 + B_2 \rho = 1 + 2^{d-1} \ph,
\label{eq:liquid_eos}
\end{equation}
where the combination of inverse temperature  $\beta $, pressure $P$, and number density $\rho=N/V$ gives a unitless quantity $p$ whose only dependence is on the liquid volume fraction $\ph=\rho V_d (\sigma/2)^d$. 

\subsection{Dynamical glass transition}

Although the structure and thermodynamics of the liquid are trivial, its dynamics is not. 
We will focus here on the equilibrium dynamics, i.e. {\it starting from an equilibrium initial condition}~\footnote{
If one considers instead the dynamics starting from an out-of-equilibrium initial condition, then the system is able to equilibrate
for $\f < \phid$ while {\it aging} effects, which persist for infinite times, are observed for $\f > \phid$~\cite{CK93,CC05}.
}.
In infinite dimension, a dynamical glass transition $\phid$ separates two distinct dynamical regimes.
For $\f < \phid$, the dynamics is diffusive at long times, as expected of any liquid.
Upon approaching $\phid$, however,
the dynamics grows increasingly sluggish, and
above $\phid$, each particle is fully confined within a cage formed by its neighbors. The typical size of that cage is the cage order parameter $\Delta_1$
(the meaning of the suffix will become clear below),
which, in that regime, can be extracted from the long-time limit of the mean-squared displacement (MSD) 
\begin{equation}\label{eq:delta}
\begin{split}
\Delta (t)&=  \frac{1}{N} \sum _{i=1}^{N} \av{|\vr_i(t) - \vr_i(0)|^2} \ , \\
\Delta_1 &= \lim_{t\to\io , \ \f > \phid} \Delta(t) \ .
\end{split}
\end{equation}

\begin{figure}[t]
\centerline{\hbox{\includegraphics [width = 0.85\columnwidth] {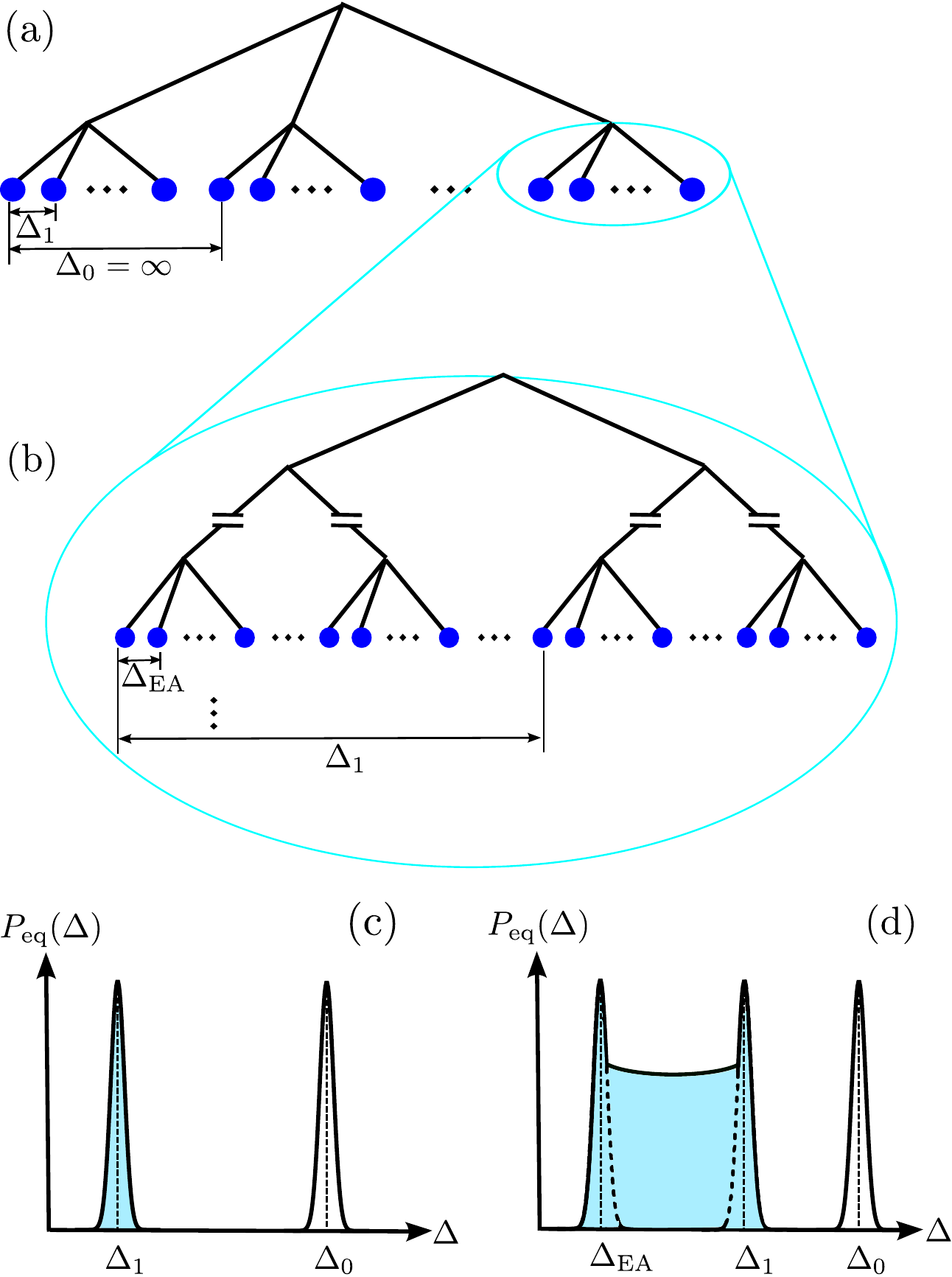}}}
\caption{(Color online). (a) Organization of glass states (blue dots) for $\phiin>\phid$; the typical intra-basin MSD 
is $\Delta_1$, while the inter-basin MSD is $\Delta_0 = \infty$. 
(b) In the SF-fullRSB phase, $\phiG < \f < \phij$, glass {metabasins subdivide} into a hierarchical structure of sub-basins with typical innermost MSD 
$\Delta_{\rm EA}$ and outermost {(meta)basin} MSD $\Delta_1$. {Schematics of the equilibrium $P_{\rm eq}(\Delta)$ and  restricted equilibrium $P_{\rm SF}(\Delta)$ (blue area) distributions are given for (c) the SF-RS (or 1RSB) and (d) the fullRSB phases.}}
\label{fig:fractal}
\end{figure}

From the $d\to\io$ solution, we know that the equilibrium distribution of the order parameter, $P_{\rm eq}(\Delta)$, has two peaks for $\f>\phid$ (see Fig.~\ref{fig:fractal})~\cite{FP95,CC05}. The first characterizes the distance between two glass configurations within a same {metabasin}. It is centered around $\Delta_1$, which is the typical size of this basin. The second characterizes the inter-basin distance. It is centered around $\Delta_0 = \infty$, because states that belong to different metabasins are completely uncorrelated. 
In technical terms, this situation is described by a 1-step {replica symmetry breaking} (1RSB) scheme~\cite{MPV87}.
Note, however, that the peak at $\D_1$ has an exponentially small weight in $N$, because there exists an exponentially large number of distinct glass 
states~\cite{CC05}. Hence, in the thermodynamic limit, $P_{\rm eq}(\Delta) = \d(\D - \D_0)$ everywhere in the liquid phase, i.e. even for $\f > \phid$.

\subsection{Glass state following and the Gardner transition}
\label{sec:gardner}

In $d\to\io$, each equilibrium configuration at density $\phiin > \phid$ is {forever} trapped into one of {the} exponentially many glass metabasins. 
The pressure of an equilibrium configuration at $\phiin$ is given by 
the liquid EOS, Eq.~\eqref{eq:liquid_eos}, but if one compresses (or decompresses) such a configuration up (or down) to a density
$\f$, the system remains %trapped 
within the metabasin that was initially selected. {The system} thus falls out of equilibrium in the sense that it
cannot visit the ensemble of all possible distinct glass metabasins and thus follows an EOS different from {that of} the liquid. % one, Eq.~\eqref{eq:liquid_eos}.

{The infinite lifetime of these $d\rightarrow\infty$ glass states implies that}
%Because in $d\to\infty$ glass states have an infinite lifetime, 
one can 
adiabatically follow the EOS of a single glass state through a {\it restricted equilibrium} approach, 
a construction also {known as} %goes under the name of 
state 
following (SF)~\cite{Franz1995,barrat1997,ZK10,Rainone2014}.
%The construction 
SF consists in using a first configuration {that is equilibrated} %at equilibrium 
at initial density $\phiin$, and a second configuration that
is in a restricted equilibrium at a different density $\f$~\cite{Franz1995,barrat1997,ZK10,Rainone2014}. {The restriction is that that second configuration must be part of the same metabasin as the first one.} %restricted because it is constrained to remain into the same glass basin, as described in. 
The state of the second configuration {thus} describes the evolution of the
glass {metabasin with  $\f$}. %, in a restricted equilibrium at density $\f$. 
Dynamically, this {process} corresponds to preparing a system at equilibrium at initial density $\phiin$,
then compressing it at density $\f$, and {\it assuming that it is able to equilibrate inside the glass {metabasin, but} without escaping it}. {In simpler words,}
%By definition thus, 
structural relaxations are frozen %during this process 
and particles {only} rattle inside their cages.
Upon compression, these cages shrink until the jamming density, $\phij(\phiin)$, is reached. 
Particles are then mechanically in contact, which makes the system mechanically rigid. 
In the following, we call the restricted equilibrium EOS of a given glass metabasin simply its glass EOS.

Examples of {different} $d\to\io$ glass EOS obtained by SF are given in Fig.~\ref{fig:theory}~\cite{Rainone2014,Rainone2666}.
In $d\to\io$, a compressed state under restricted equilibrium undergoes a Gardner transition at $\phiG(\phiin)$, 
at which point the glass metabasin {subdivides} into a hierarchy of sub-basins. 
Technically,
before the Gardner transition, i.e., for $\f < \phiG(\phiin)$, the glass {metabasin} is {obtained} by a replica symmetric SF (SF-RS) computation~\footnote{A replica symmetric SF computation is {quite} similar to a standard 1RSB computation.}, while
in the Gardner phase, i.e., for $\f > \phiG(\phiin)$,
a full replica symmetry breaking SF (SF-fullRSB) {computation} is needed~\cite{Ga85,GKS85,MPV87,CKPUZ14,Charbonneau2014B,Rainone2014}. 
The Gardner transition is therefore akin to the {spin glass transition} one 
of the SK model, but {restricted} %within the restriction 
to a {single} glass metabasin.
The high temperature phase of the SK model {thus} corresponds to a simple glass metabasin for $\f < \phiG(\phiin)$ while the low temperature phase
corresponds to a fractured and marginal metabasin for $\f > \phiG(\phiin)$.
%For the purpose of this work, 
{Note that we here complement} the SF-RS computation of Ref.~\cite{Rainone2014} with a SF-fullRSB computation, 
{in order} to obtain
the complete EOS reported in Fig.~\ref{fig:theory}. The SF-fullRSB equations are similar to the ones reported in Ref.~\cite{Charbonneau2014B}, {but the full}
details {of this work are} given elsewhere~\cite{Rainone2666}.

Following a {glass} state in restricted equilibrium gives
for $\f < \phiG(\phiin)$ a distribution $P_{\rm SF}(\Delta)$ that has a single peak at $\Delta_1$. 
Note that because the system is confined to a single glass basin, the peak at $\D_0$ is absent.
At $\phiG(\phiin)$ the glass basin fractures in a SF-fullRSB structure, and correspondingly
the single peak in the distribution $P_{\rm SF}(\Delta)$ fractures into two peaks centered around
$\Delta_{\rm EA}$ and $\Delta_1$, connected by a wide continuous band (see Fig.~\ref{fig:fractal} and~\cite{MPV87}). 
Here, $\Delta_{\rm EA}$ is the typical size of the innermost sub-basins at the lowest hierarchical level, 
and $\Delta_{\rm 1}$ is the typical {distance between} %size of 
the outermost {sub-basins, i.e., the size of the metabasin}.  
For the same reason as above, $P_{\rm SF}(\Delta)$ does not show the $\D_0$ peak.

Because the cage order parameter changes continuously at $\phiG$, the Gardner transition is a continuous critical transition~\cite{Ga85,GKS85}. 
The Gardner phase is also marginally stable, in the sense that a zero mode is always present in the stability matrix of the free energy~\cite{MPV87}.
Because jamming is located within the Gardner phase, its marginal stability and critical scaling behaviors can consequently be obtained from a 
fullRSB thermodynamic calculation~\cite{CKPUZ14,Charbonneau2014B}.

\subsection{Timescales}
\label{sec:timescales}

%We now discuss 
The timescales that characterize the dynamics in the different phases of the MK model are predicted based on the general correspondence between
statics and dynamics in spin glasses~\cite{CK93,CK94}. Here {again, we only consider the system behavior in the} limit $d\to\io$.
%In addition to 
{Beyond} the microscopic timescale $\t_0$, over which dynamics is essentially ballistic, one gets the following picture.
\begin{itemize}
\item
In the liquid phase below $\f_{\rm d}$, dynamics is characterized by two timescales: the $\b$-relaxation timescale $\t_\b$, over which particles
explore their {transient cage}, and a longer timescale $\t_\a$, over which dynamics is diffusive. 
Both timescales are finite for $\f < \f_{\rm d}$, and diverge at $\f_{\rm d}$ according to the scaling
predicted by mode-coupling theory (MCT)~\cite{Go09}. {We do not discuss this regime further because it is not directly related to the Gardner transition}.
\item
In the liquid phase above $\f_{\rm d}$ and {in the SF-RS (simple glass) phase}, the same two timescales exist:
$\t_\a \sim \exp(N)$ is the timescale for jumping from one glass {metabasin} to another, which {being infinite in the thermodynamic limit properly defines the glass metabasins,} and
%hence basins are well defined.
$\t_\b$ is the timescale for equilibrating {within} a glass {metabasin, i.e., for particles to explore their cage,} which is finite for $\f > \phid$ %. Note that $\t_\b$ also 
{but diverges}
upon approaching $\phid$ from densities above it, again following a MCT-like scaling form~\cite{Go09}.
\item
Upon approaching the Gardner transition, $\t_{\b}$ % that is needed to equilibrate inside the glass basin
again diverges, {scaling as} $\t_{\b} \sim (\phiG - \f)^\g$.
%exactly for 
{It does so for the exact} same reason as {at near the spin-glass transition of} the SK model %where the equilibration time diverges at the phase transition: in both cases the divergence of the relaxation time is due to the 
{and corresponds to the} critical slowing down close to a second-order
phase transition.
\item
In the Gardner phase, {dynamics is described by three timescales}:
$\t_\b$ is the timescale for equilibrating {within} a single glass sub-basin ($\D < \D_{\rm EA}$),
{$\t_{\rm meta}$ is the timescale over which system explores the structure of the sub-basins within a given glass metabasin
($\D_{\rm EA} < \D < \D_1$), and $\t_\a \sim \exp(N)$ remains the timescale for jumping from one glass metabasin to another ($\D \sim \D_0$)}. To be more {clear}, none of these processes correspond to a simple exponential with a single timescale.
{Everywhere in the Gardner phase $\t_\b=\infty$, because of the phase's marginality. 
The relaxation inside a single sub-basin is thus expected to scale as a power-law in time.} 
The exploration of sub-basins is
characterized
by a complex distribution of free energy barriers and relaxation times, and {thus} $\t_{\rm meta} \sim \exp(N^\a)$ with $\a<1$
({$\a=1/3$} is expected in the SK model~\cite{MPV87}). Note that 
barriers between sub-basins are much lower than {those} between metabasins, hence $\t_{\rm meta} \ll \t_\a$.
\end{itemize}

Let us now situate {the scheme for} our work in this complex $d\rightarrow\infty$ dynamic phase diagram. We {are conveniently} {\it not} concerned with the scalings
of the MCT regime. %In fact, 
Thanks to the planting procedure (see Sec.~\ref{sec:planting}), we are {indeed} able to {easily} generate
{\it equilibrium} configurations {of the MK model for arbitrary $\phiin$, including $\phiin > \phid$}. These configurations are well equilibrated in a glass {metabasins} and therefore
have an infinite $\t_\a$ (in the thermodynamic limit) and a finite $\t_\b$. {Because $\t_\a=\infty$ in this density regime,} we can simply forget
about its existence and consider that we are forever restricted into {a given} glass {metabasin}. From this point of view, we are in
a situation similar to that of a spin glass
where one is able to start at equilibrium in the paramagnetic, high temperature phase.

If we compress the system slowly enough to a final density $\f \in (\phiin, \phiG)$, {then equilibration within
the glass {metabasin} is possible, because $\t_\b$ remains finite. In this situation, the system {behavior} is described by the SF-RS {computation}.
%This is possible if we are not too close to $\phiG$, {such that the simulation timescale} remains larger than $\t_\b$.
To study the Gardner transition, however, we need to compress the system up to $\phiG$, at which $\t_\b$ diverges.
Hence, upon approaching $\phiG$,  $\t_\b$ eventually becomes larger than the simulation timescale}, and equilibration (even in the restricted SF
sense) becomes impossible. The system {thus} falls out of (restricted) equilibrium and is not described anymore by the SF computation. 

For $\f > \phiG$ the situation is even worse {both} as $\t_\b$ and $\t_{\rm meta}$ are infinite. The SF-fullRSB computation, which gives
the restricted equilibrium properties in this regime, is only an approximation even at long (but not divergent with {$\exp(N)$}) times. {The situation is akin to that}
in the SK model where the fullRSB computation is only
an approximation of the states reached dynamically at long times in the spin-glass phase.  
{Because} in this regime the planting technique does not work (it only works on the liquid line in Fig.~\ref{fig:theory}), we cannot use it
to study the restricted equilibrium in the Gardner phase.

In the following we will {therefore} present two kinds of data:
\begin{itemize}
\item for $\phiin \leq \f < \phiG$, far enough from $\phiG$ such that $\t_\b$ is smaller than the simulation timescale, we %will have access to 
{obtain} restricted 
equilibrium data
\item for $\f \sim \phiG$ and $\f > \phiG$, the system is out of (restricted) equilibrium, shows {\it aging} effects, and, at long times, {we obtain states
that are qualitatively similar to the SF-fullRSB ones, but not exactly equal to them}. %, like in the SK model at low temperatures.
\end{itemize}
In Sec.~\ref{sec:numerics} we give a more precise definition of the numerical protocol we use to study these different regimes.
Note that, according to the above discussion, from now on we will refer to the restricted equilibrium simply as ``equilibrium", given that
we always work inside a glass {metabasin}.

\subsection{Finite-dimensional MK model}
\label{sec:finiteD}

To conclude this section, we discuss the additional effects that appear when one considers the MK model in finite dimensions, and that affect
the above discussion.

%First of all, f
From a theoretical point of view, a finite $d$ affects quantitatively the phase diagram, but to a lowest degree of approximation 
one can simply
take the phase diagram of Fig.~\ref{fig:theory} {and fix the value of $d$ (e.g. $d=3$) to obtain result for the EOS}. However, there are
several systematic corrections that impact the accuracy of this result.
\begin{itemize}
\item
The infinite-dimensional results are obtained within a Gaussian structure of the cage, 
which provides the exact results in the limit $d\to\io$~\cite{KPZ12}. The Gaussian equations
are slightly different in finite $d$~\cite{PZ10}. The corrections have the form of a series in $1/d$ and
are quite small. For example, in $d=3$ one has {$2^3 \phid/3 \approx 4.74$} instead of the infinite-dimensional
result $2^d \phid/d \approx 4.8$. These corrections are negligible and could be easily taken into account if needed.
\item 
A more important problem is {the inexactitude} of the Gaussian assumption for the MK model in finite dimensions. One should {instead}
optimize the free energy over a generic cage function, {but this computation is technically quite difficult}. In Ref.~\cite{Charbonneau2014}, {however,} two different {cage functions
%for the cage have been 
were studied and found to give similar results}. The corrections coming from the non-Gaussianity of the cage
are {thus} also rather small, {although they are} more difficult to estimate.
\item
In finite $d$, cages are heterogeneous~\cite{Charbonneau2014}. {Carefully taking this effect into account} would require a cavity calculation
similar to the one performed in Ref.~\cite{Charbonneau2014}, which is beyond the scope of this work, {but is also likely a small correction for $\phiin>\phid$~\cite{Charbonneau2014}}.
\end{itemize}
For all these reasons, the results of Fig.~\ref{fig:theory} are only an approximation to the finite-dimensional phase diagram.
%Still, 
In the following, {we will nonetheless take the results of the simplest theoretical approximation}, namely take the $d\to\io$ results 
of Fig.~\ref{fig:theory} and use them in $d=3$, {after properly rescaling the axes}.
We expect (and check) that the corrections are quite small. Sometimes, {however, in order to take effective account of} these corrections,
we will {further} rescale the {results, so as to} obtain a better {quantitative} agreement between theory and simulation. %{When that is the case, explicit mention of this slight adjustment is made in the text below}.

%Still, we cannot use the theoretical values for the Gardner transition in the numerical analysis, because even
%the small errors discussed above that affect the determination of the transition point would make the analysis impossible,
%because the interesting quantities diverge at $\phiG$.
%Instead, the values of the dynamical critical exponents obtained from theory should be reliable enough: a small error in the exponents does not
%have a deep impact on the analysis.
%For this reason we will fit the Gardner
%transition point in the numerical analysis, and then compare it to the theoretical result.

The most important genuinely finite-dimensional effect is {\it hopping}.  
As discussed in Ref.~\cite{Charbonneau2014}, slightly above $\phid$, particles are not perfectly confined in their cages,
as one would expect based on the $d\to\io$ picture. Instead, each particle is allowed to explore a network of cages,
connected by narrow pathways. Hopping consists in particles jumping between distinct cages. The corresponding time
scale is finite at $\phid$ and thus the $d\to\io$ divergence of the relaxation time is washed away in finite $d$. 
Hopping effects also change deeply the dynamics of the system with respect to the $d\to\io$ prediction, and
the configurations at 
$\phiin \gtrsim \phid$ are not able to constrain the dynamics for infinitely-long times. {Because the timescale for hopping increases quickly upon increasing both density and dimension, considering values of $\phiin$ that are slightly above $\phid$ suffices}.
In practice, {based on the analysis of} Ref.~\cite{Charbonneau2014}, in $d=3$ and
for $\phiin \geq 2.5$ hopping is strongly suppressed, hence in that regime the mean-field, {$d\rightarrow\infty$} scenario should apply {reasonably} well.

\section{Numerical Approach}
\label{sec:numerics}
In this section, we provide the numerical details used in the simulations of the glass states of the MK model.

\subsection{Planting}
\label{sec:planting}
An important algorithmic advantage of the MK model is that planting can be used to generate equilibrium liquid configurations at any $\phiin$~\cite{KZ09,Charbonneau2014}. This procedure sidesteps the tedious and time-consuming work of first preparing dense equilibrium configurations, as would be needed for typical glass formers, such as HS. The basic idea is to switch the order in which  initial particle positions $\{\br_i\}$ and random shifts $\{\mathbf{\Lambda}_{ij}\}$ are determined. 
One first chooses the particle positions $\{\br_i\}$ independently and uniformly in the volume $V$, and then for each particle pair one chooses
a random shift $\mathbf{\Lambda}_{ij}$ uniformly under the sole constraint that the two particles should not overlap, which is quite straightforward to satisfy.
As long as the quenched and the annealed averages of the free energy are the same
(see Ref.~\cite{KZ09} for a more detailed discussion), 
a planted state is a true equilibrium state and
automatically satisfies the liquid EOS, Eq.~\eqref{eq:liquid_eos}. This condition is met along the replica symmetric phase for $\phiin < \f_{\rm K}$, where $\f_{\rm K}$ is the Kauzmann point at which the configurational entropy vanishes~\cite{PZ10,MK11}. Because in the MK model $\f_{\rm K} = \infty$~\cite{MK11}, planting a liquid configuration is thus possible at any density, which dramatically reduces the computational cost of the initial equilibration.

In our notation, a given
$\{\br_i\}$ and $\{\boldsymbol{\Lambda}_{ij}\}$ defines a {\em sample}.
A sample thus identifies a given system (defined by
$\{\boldsymbol{\Lambda}_{ij}\}$) and, for this system, one of its glass {metabasins} %states 
(selected by $\{\br_i\}$).

\subsection{Molecular dynamics (MD) simulations}

We use event-driven molecular dynamics (MD) to
simulate MK particles in $d=3$~\cite{SDST06,Charbonneau2014}. Periodic boundary conditions with the minimum
image convention are implemented on the shifted distances $|\br_i - \br_j +
\boldsymbol{\Lambda}_{ij}|$. Time $t$ is expressed in units of $\sqrt{\beta m
  \sigma^2}$, where the particle mass $m$ and diameter $\sigma$ as well as the
inverse temperature $\beta$ are set to unity. Systems consist of $N= 800$
particles unless otherwise specified. This system size is large enough to
contain a first full shell of neighbors around each particle, and to keep the
periodic boundary effects on caging to a minimum~\cite{Charbonneau2014}.
Finite-size effects are studied for some of the observables for the initial
liquid density $\phiin= 2.5$ (see Sec.~\ref{sec:static}).

To simulate SF, for a given sample, we start from the planted equilibrium configuration  at a
packing fraction $\phiin$, and grow the spheres following the
Lubachevsky-Stillinger algorithm~\cite{DTS05,SDST06} at constant growth rate
$\gamma = 0.001$, unless otherwise specified, up to a desired $\f$. 
Once
compression is stopped at the target density, the origin of time is set. {We will thus typically (although not always) define 
the waiting time $t_w$, as the time that has elapsed since the end of the compression}.
From that moment on
we start measuring observables, keeping
density and temperature (and thus energy) constant. 
Note that because $\gamma$ is finite and rather small, part of the equilibration happens already during compression,
so provided we are not too close to $\phiG$ the system is stationary at all $t_w$ (we come back on this point later).
This procedure
is repeated over $N_\mathrm{s}$ samples in order to average over thermal and
quenched disorders. Errors are computed using the jack-knife method~\cite{amit2005field}. Depending on the statistical convergence of the different observables, $N_{\mathrm{s}}$ is varied from 500 to 75,000, as specified in the discussion of the various measurements.

\subsection{Observables}
\begin{figure}[t]
\centerline{\hbox{\includegraphics [width = \columnwidth] {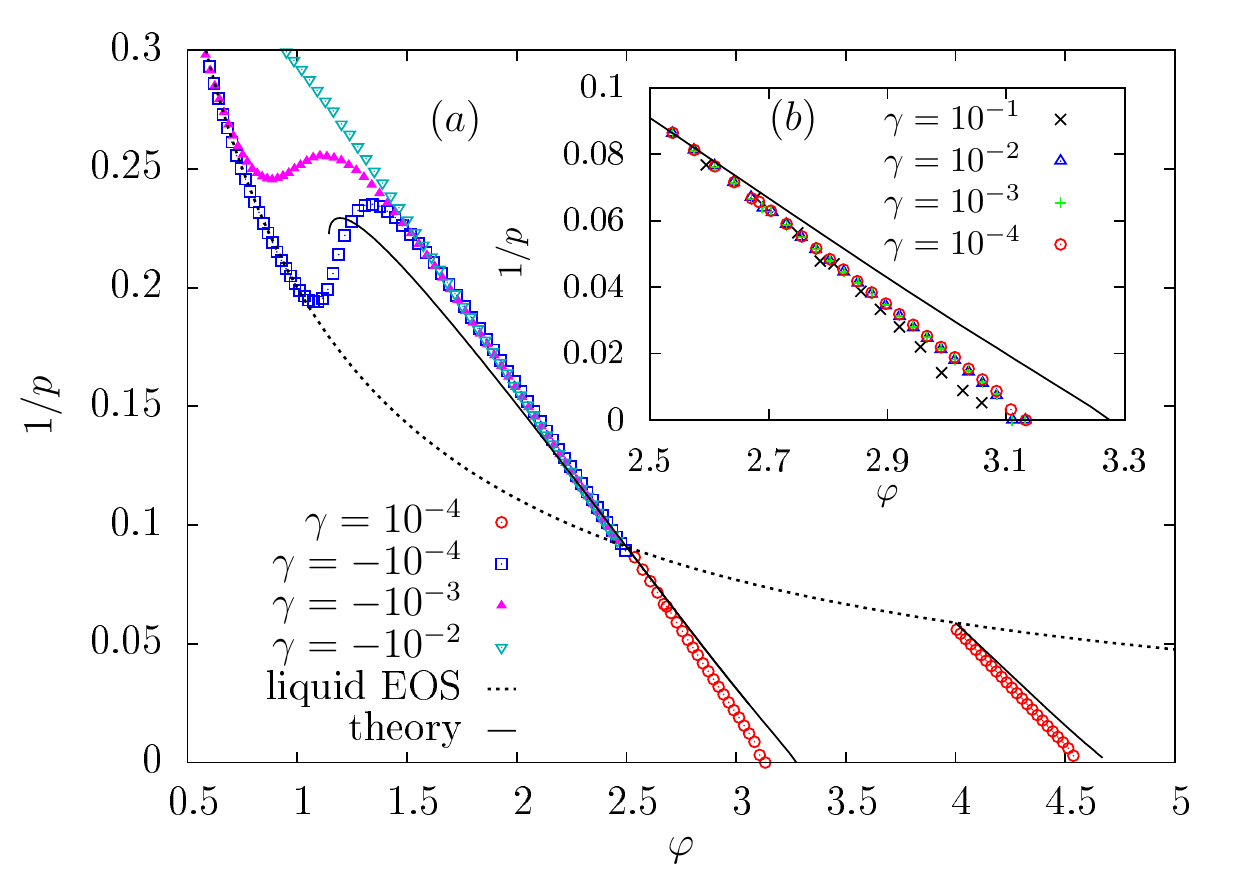}}}
\caption{(Color online). (a) 
Compressions ($\gamma>0$) and decompressions ($\gamma<0$) of an initial equilibrium state at $\phiin = 2.5$ and 4.0. The results are averaged over $N_{\mathrm{s}}=100$ samples. The theoretical curves uses data from Ref.~\cite{Rainone2014} and from this work~\cite{Rainone2666}. 
%On the scale of this figure, results for $\gamma \leq 0.001$ would be indistinguishable, hence only $\gamma=0.001$ is shown. 
(b)~Compression data are indistinguishable for $\gamma \leq 0.01$.
}
\label{fig:state_following}
\end{figure}

The pressure evolution along SF is reasonably well described by a free-volume EOS
\begin{equation}
\frac{1}{p} = C \left( 1 - \frac{\ph}{\phij}\right),
\label{eq:free_volume}
\end{equation}
where $C$ is a fitting parameter. Fitting Eq.~\eqref{eq:free_volume} to the compression results provides an estimate of $\phij$ (see Table~\ref{tab:phiG}). 
If a sufficiently small $\gamma$ is chosen, no aging is observed in the pressure, and  
using slower compression rates gives only negligible corrections to the glass EOS (see Fig.~\ref{fig:state_following}). 
Interestingly, upon decompression, the state follows the same EOS up to a threshold density at which it melts into a liquid phase. This phenomenon has been recently predicted by the theory~\cite{MPR14,Rainone2014}, and observed numerically in simulated ultrastable glasses~\cite{Singh2013, Hocky2014}. 

To obtain more structural information about the free energy landscape, we also simulate a {\it cloning} procedure. The approach consists of taking two exact copies (clones) $A$ and $B$ of the same planted configuration at $\phiin$, and assigning them different initial velocities, randomly drawn from the Maxwell--Boltzmann distribution. These two copies are then independently compressed up to  $\f$, before measuring the mean-squared distance between them
\begin{equation}\label{eq:deltaAB}
\DAB (t)=  \frac{1}{N} \sum _{i=1}^{N} \av{|\vr_i^A(t) - \vr_i^B(t)|^2},
\end{equation}
where $\vr_i^A(t)$ and $\vr_i^B(t)$ are the positions of particle $i$ at time $t$ in clones $A$ and $B$, respectively.
Although the two clones start from the same initial configuration, {their compression histories are different} once $\f$ is reached.

The detailed behavior of the clones will be discussed in Sec.~\ref{sec:time}, 
but let us explain here briefly why the cloning procedure is useful to detect
the Gardner transition.
In the SF-RS phase, 
the two clones are uncorrelated in the glass basin and $\DAB(t)$ converges quickly
(on a time scale $t \sim \t_\b$ if the two clones are
not sufficiently well equilibrated along the compression)
to the equilibrium value $\DAB = \D_1$. Hence $\DAB = \lim_{t\to\io} \D(t)$ in the SF-RS phase.
{By contrast,} if the end point of the compression falls {within} the {SF-fullRSB} (or Gardner) phase, clones 
most likely fall into different sub-basins. Their mean square distance can then be described by a non-trivial 
time-dependent probability distribution, $P_{AB}(t,\D)$,
that depends on the way sub-basins are sampled. Calculating these weights is difficult, because the two clones are generally out of equilibrium.
Because the probability that the two clones fall in the same state is very small, {however,} and therefore
%$\DAB(t)$ is, at all times, such that 
$\DAB(t) > \DEA$ for all $t$. {Hence, in the Gardner
phase $\DAB(t)$ at all $t$ is strictly larger than the long-time limit of the MSD.}
%Thus, 
The long time limit of $\DAB - \D$, {being} zero in the SF-RS phase {and} non-zero in the SF-fullRSB phase,
thus provides an {\it order parameter} for the Gardner transition.

\section{Detecting the Gardner transition}
\label{sec:GardnerDetection}

In this section, we describe different means of detecting the Gardner transition through numerical simulations designed to follow the evolution
of glassy states. Let us stress once again that the aim of this work is to understand, in a controlled setting, how well numerical simulations and experiments 
can detect the Gardner
transition. {We can} therefore test different strategies to understand their advantages and limitations, and use the analytical results to assess the quality
of the numerical results.
Based on the discussion in Sec.~\ref{sec:II}, 
we follow two complementary approaches. The first is based on dynamics. From the long-time dependence of the mean square displacement, 
we determine {$\t_\b$}, %the characteristic relaxation time 
whose divergence in the glass {metabasin} signals the Gardner transition. 
The second is based on the properties of the distribution function $P_{AB}(\D)$, which becomes non-trivial in the Gardner phase. We investigate
different moments of this distribution to {detect} signatures of the transition.

\subsection{Dynamics}
\label{sec:time}

As discussed in Sec.~\ref{sec:timescales}, the Gardner transition is a second order phase transition 
associated with a diverging characteristic relaxation time $\t_\b$ that controls the dynamics of the MSD.
To detect this relaxation time we make use of the following observables already briefly discussed
in Sec.~\ref{sec:numerics}. Recall that the origin of time $t=t_w=0$ is set at the end of the compression when
a given final density $\f$ {is} reached.

\subsubsection{Definition of the relevant observables}
\label{sec:IVA1}
We define the MSD between two configurations at
different times:
\begin{equation}
\begin{split}
\widehat\Delta (t,t_w) =  \frac{1}{N} \sum _{i=1}^{N} |\vr_i(t + t_w) - \vr_i(t_w)|^2 \ , 
\end{split}
\end{equation}
and the MSD between two different clones
\begin{equation}
\begin{split}
\widehat\Delta_{AB} (t) =  \frac{1}{N} \sum _{i=1}^{N} |\vr^A_i(t) - \vr^B_i(t)|^2 \ .
\end{split}
\end{equation}
From these two instantaneous quantities we can define different observables. The statistical
average (over compressions and samples) gives
\beq
\D(t,t_w) = \av{
\widehat\Delta (t,t_w)
} \ ,
\hskip20pt
\DAB(t) =\av{
\widehat\Delta_{AB}(t)
} \ ,
\eeq
{and we define}
%For later use, we also define a quantity
\beq
\d \D(t,t_w) = \DAB(t + t_w) - \D(t,t_w) \ .
\eeq
We also define a time-dependent caging susceptibility as the normalized variance of the MSD
\begin{equation}\label{eq:chi1}
\chi(t,t_w) =N\frac{\av{\widehat\D^2(t,t_w)}-{\av{\widehat\D(t,t_w)}^2}}{\av{\widehat\D(t,t_w)}^2},
\end{equation}
and its counterpart of cloned configurations 
\begin{equation}\label{eq:chi2}
\chi_{AB}(t) =N\frac{\av{\widehat\D_{AB}^2(t)}-{\av{\widehat\D_{AB}(t)}^2}}{\av{\widehat\D_{AB}(t)}^2}.
\end{equation}

In the SF-RS phase in {restricted} equilibrium, {i.e., at large enough $t_w$ and $t> 0$,}
$\D(t,t_w) = \D(t)$ gives back Eq.~\myref{eq:delta}, 
while
$\DAB(t) = \D_1$ does not depend on time.
In this case 
\beq\label{eq:Dasy}
\D_1 = \DAB( \forall t) =  \lim_{t\to\io} \D(t)
\eeq
gives the average cage radius of the glass basin.
Similarly, $\chi(t,t_w) = \chi(t)$ and $\chi_{AB}(t) = \chi$, where 
\beq
\chi = \chi_{AB}(\forall t) =  \lim_{t\to\io} \chi(t)
\eeq
is the average susceptibility of the glass basin.
Finally, note that
based on Eq.~\eqref{eq:Dasy}, we have
\beq\label{eq:longdD}
\lim_{t\to\io} \lim_{t_w\to\io} \d\D(t,t_w) = 0 \ .
\eeq
%On the contrary, 
{By contrast,} in the SF-fullRSB phase, equilibrium is not reached even for very large $t_w$ ({recall that we do not consider here
times that are comparable to $\exp(N)$}). Therefore, $\d\D(t,t_w)$ remains non-zero even for large $t_w$ and $t$, {which makes the long time limit
of $\d\D(t,t_w)$} a dynamic order parameter for the Gardner transition.

\subsubsection{Qualitative change in caging and susceptibility}

\begin{figure}[t]
\includegraphics [width =\columnwidth] {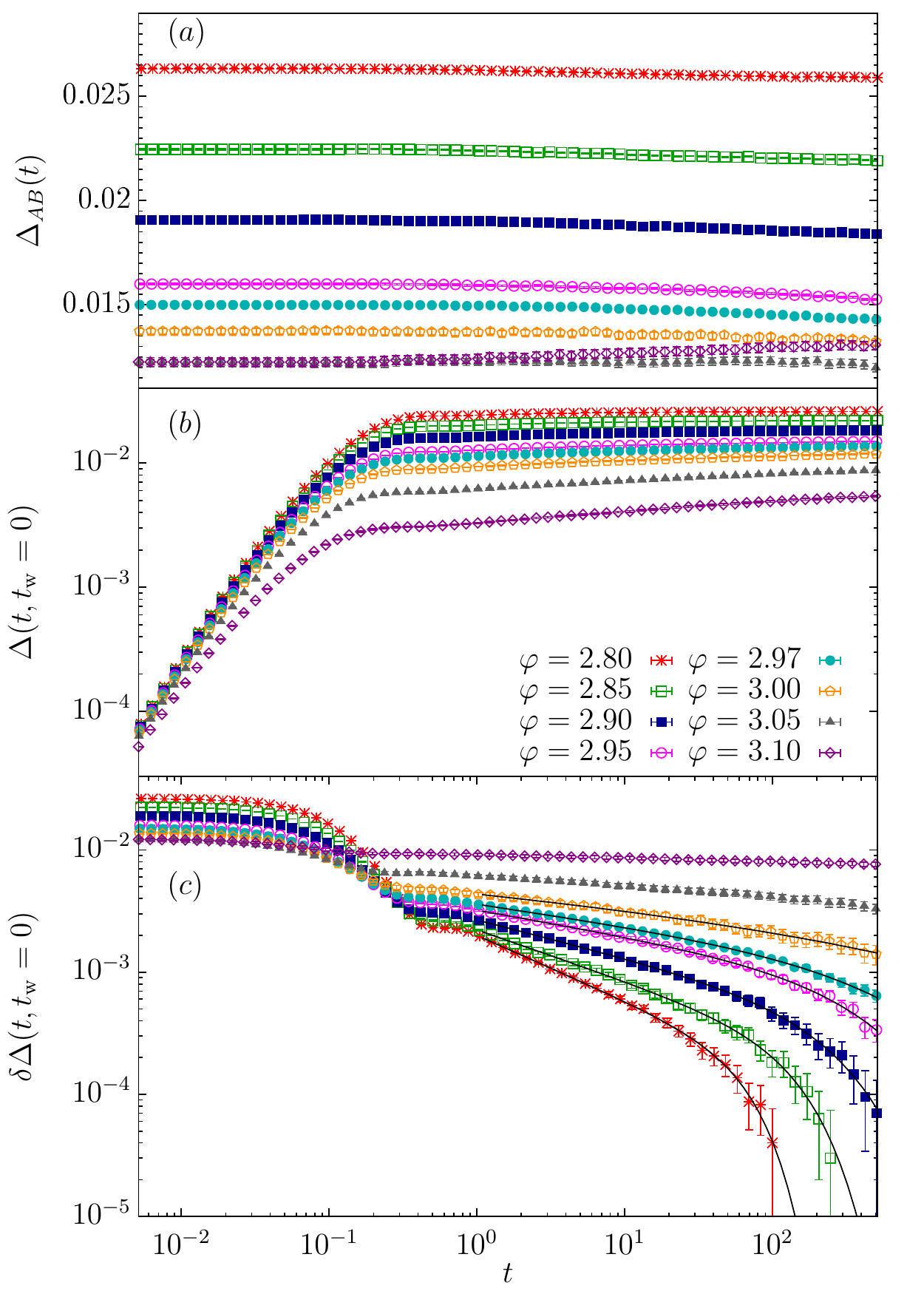}
\caption{(Color online). SF from $\phiin= 2.50$ gives 
(a) $\Delta_{AB}(t)$, (b) $\Delta(t, t_w = 0)$, and  (c) $\delta \Delta(t, t_w = 0)$ at different $\varphi$. 
Solid lines are fits to Eq.~\eqref{eq:fit_tau} (bottom panel). 
 }
\label{fig:delta-t}
\end{figure}

\begin{figure}[t]
\includegraphics [width =\columnwidth] {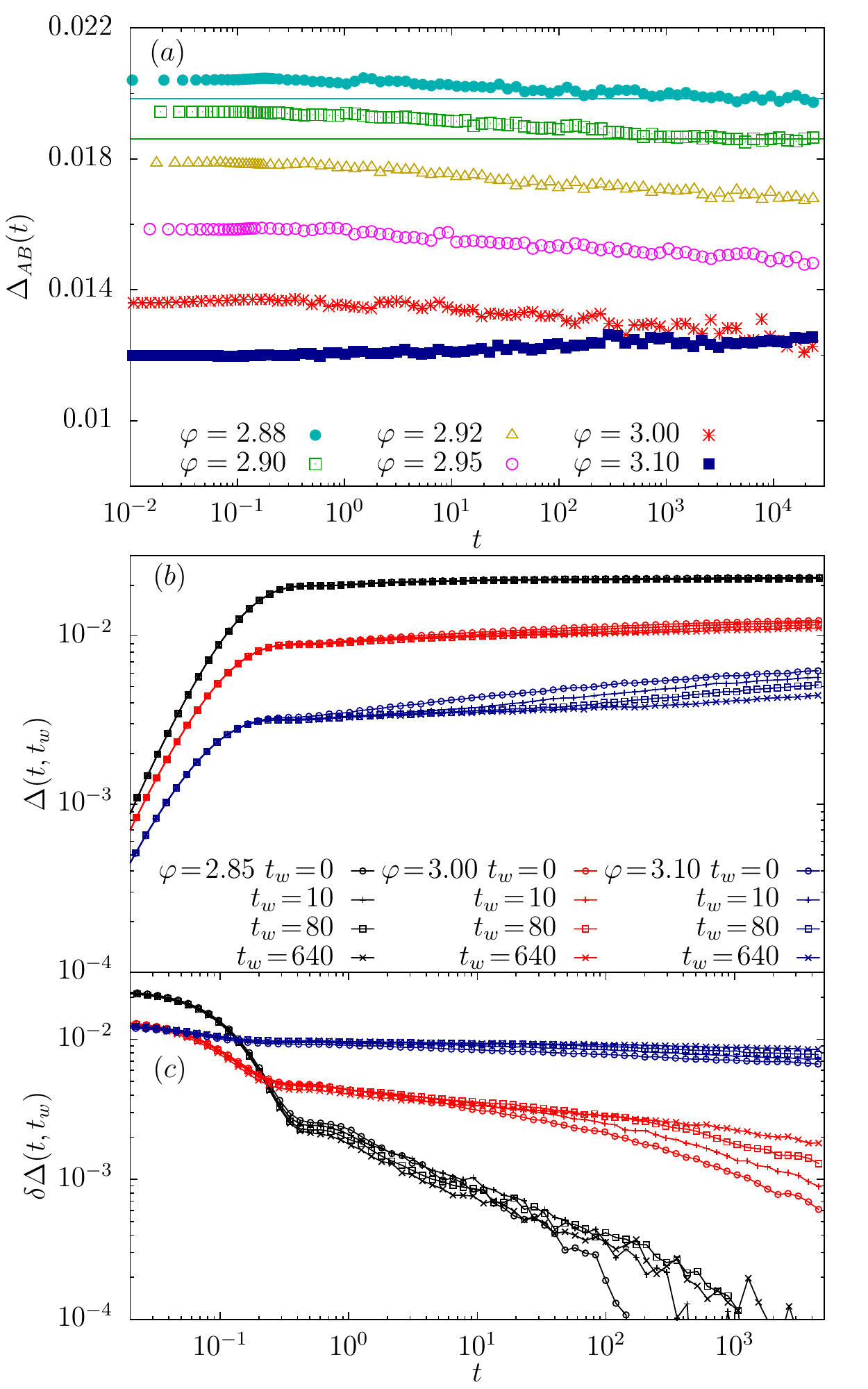}
\caption{(Color online). (a) Long-time results for $\Delta_{AB}(t)$. 
The data show that, for $\f < \phiG$, $\Delta_{AB}(t)$ saturates at its asymptotic value (dotted line) in the long-time limit (see, e.g., $\varphi = 2.88,  2.90$). (b) $\Delta(t, t_w)$ and (c)  $\delta \Delta(t, t_w)$ are
 plotted as a function of $t$ for {various} $\varphi$ and $t_w$.}
\label{fig:delta-ttw}
\end{figure}

Figure~\ref{fig:delta-t} shows the 
time dependence of 
$\DAB(t)$, $\D(t,t_w=0)$ and $\d\D(t,t_w=0)$
for $\phiin=2.5$, averaged over $N_{\mathrm{s}}=15,000$ to $N_{\mathrm{s}}=500$
 from the lowest and highest $\f$, respectively.
Both $\Delta$ and $\DAB$ are observed to behave slightly differently below and above $\phiG$
($\phiG\approx 3.00$ for $\phiin=2.50$, a more precise estimate {is obtained} below). 
For $\ph<\phiG$, $\Delta(t,t_w=0)$
first (and up to a microscopic time $\tau_0 \sim 10^{-1}$)  grows quickly because of the ballistic motion of particles~\cite{Charbonneau2014} and then more slowly in the $\b$ relaxation regime, before eventually reaching the plateau $\D = \D_1$ that defines the cage size. This plateau coincides with the (almost time-independent) results for $\DAB(t)$, as is qualitatively expected for a system in a SF-RS phase.
For $\ph \gtrsim \phiG$, the situation is a bit more convoluted. 
As discussed above, beyond the Gardner transition each of the original metabasins is expected to subdivide into a hierarchical 
distribution of glassy states.
Equilibration within the glass metabasin is now, {however,} impossible, and we thus observe that $\DAB(t)$ depends on time for all observable
times while $\D(t,t_w=0)$ remains always strictly smaller than $\DAB(t)$ and never reaches a plateau. 
Correspondingly $\d \D(t,t_w=0)$
does not decay to zero.

In Fig.~\ref{fig:delta-ttw}, we {select a few densities and consider} the evolution of $\DAB(t)$ over much larger times than in Fig.~\ref{fig:delta-t} 
{as well as} the dependence of $\Delta(t,t_w)$ on both $t$ and $t_w$. %, for selected densities; 
{Note that} these data are averaged over {many fewer} %smaller set of 
samples, $N_{\rm s} \approx 300$.
 {We consider the detailed behavior of these two quantities}: %Let us describe in detail the behavior of these quantities:
\begin{itemize}
\item
{In general,} $\DAB(t)$ (Fig.~\ref{fig:delta-ttw}a)
should correspond to the average distance between {configurations restricted to} within a {given} metabasin, 
but for {$t<\t_\b$} %at short times 
the basins are sampled with non-equilibrium weights, and
hence $\DAB(t)$ {slowly} drifts. % on a slow timescale. 
For $\f < \phiG$, we {indeed observe} that {once} $t \sim \t_\b$, 
$\DAB(t)$ reaches
a stationary value. Upon approaching $\phiG$ and for $\f > \phiG$, {however,} $\t_\b$ is {becomes so large that} %infinite and 
the drift of $\DAB(t)$ persists at all {simulated} times.
Note that the drift {can be positive or negative}, %towards decreasing or increasing values of $\DAB$ 
depending on density.

\item
The behavior of $\D(t,t_w)$ (Fig.~\ref{fig:delta-ttw}b) is naturally described as a function of $t$ for fixed $t_w$.
Below $\phiG$ (e.g., for $\f=2.85$), $\D(t,t_w)$ is independent of $t_w$ and behaves as in Fig.~\ref{fig:delta-t}.
Beyond $\phiG$ (e.g., for $\f=3.10$), {however,}
the system is initially %(immediately after stopping the compression) 
trapped into a sub-basin, 
and thus $\Delta(t, t_w)$
grows until reaching a plateau corresponding to the size of the sub-basin, $\DEA$, for $t \ll \t_{\rm meta}(t_w)$. 
For $t \sim \t_{\rm meta}(t_w)$, the system {can} explore the structure of sub-basins, hence
$\Delta(t, t_w)$ keeps increasing, 
and no clear first plateau can be detected. 
A second plateau should be reached in the limit $t  \gg \t_{\rm meta}(t_w)$, when the metabasin is fully explored,
but this regime is here beyond computational reach.
The timescale $\t_{\rm meta}(t_w)$ {indeed} %is %initially short and 
increases with $t_w$~\cite{CK93,CK94}, {as} is clearly visible in
Fig.~\ref{fig:delta-ttw}, {where} the drift shifts towards larger times upon increasing $t_w$.
Note that in the limit $N\to\io$, $\t_{\rm meta}(t_w)$ should
diverge for $t_w\to\io$, {but} a finite $N$ acts as a cutoff and $\t_{\rm meta}(t_w)$ should {instead} saturate to a value $\sim \exp(N^{1/3})$ for
large $t_w$.

\item
%Finally, we observe that 
In the Gardner phase $\D(t,t_w) < \DAB(t)$ for all accessible times $t,t_w$.
Therefore, {although} $\d\D(t,t_w)$ goes to zero at large times for $\f < \phiG$, for
$\f \gtrsim \phiG$ it does not {fully} decay in the accessible $t$ regime (for all $t_w$) and instead converges
to a plateau (Fig.~\ref{fig:delta-ttw}c). Note also that the evolution of 
$\d\D(t,t_w)$ with increasing density is qualitatively identical to {that} of the overlap with decreasing temperature
in spin glasses, as reported in Ref.~\cite[Fig.~8]{Ogielski1985} ({compare also} to Fig.~\ref{fig:delta-t}c). 
{These results further support the strong analogy between the Gardner and the spin-glass transitions}.

\end{itemize}

\begin{figure}[t]
\includegraphics [width =\columnwidth] {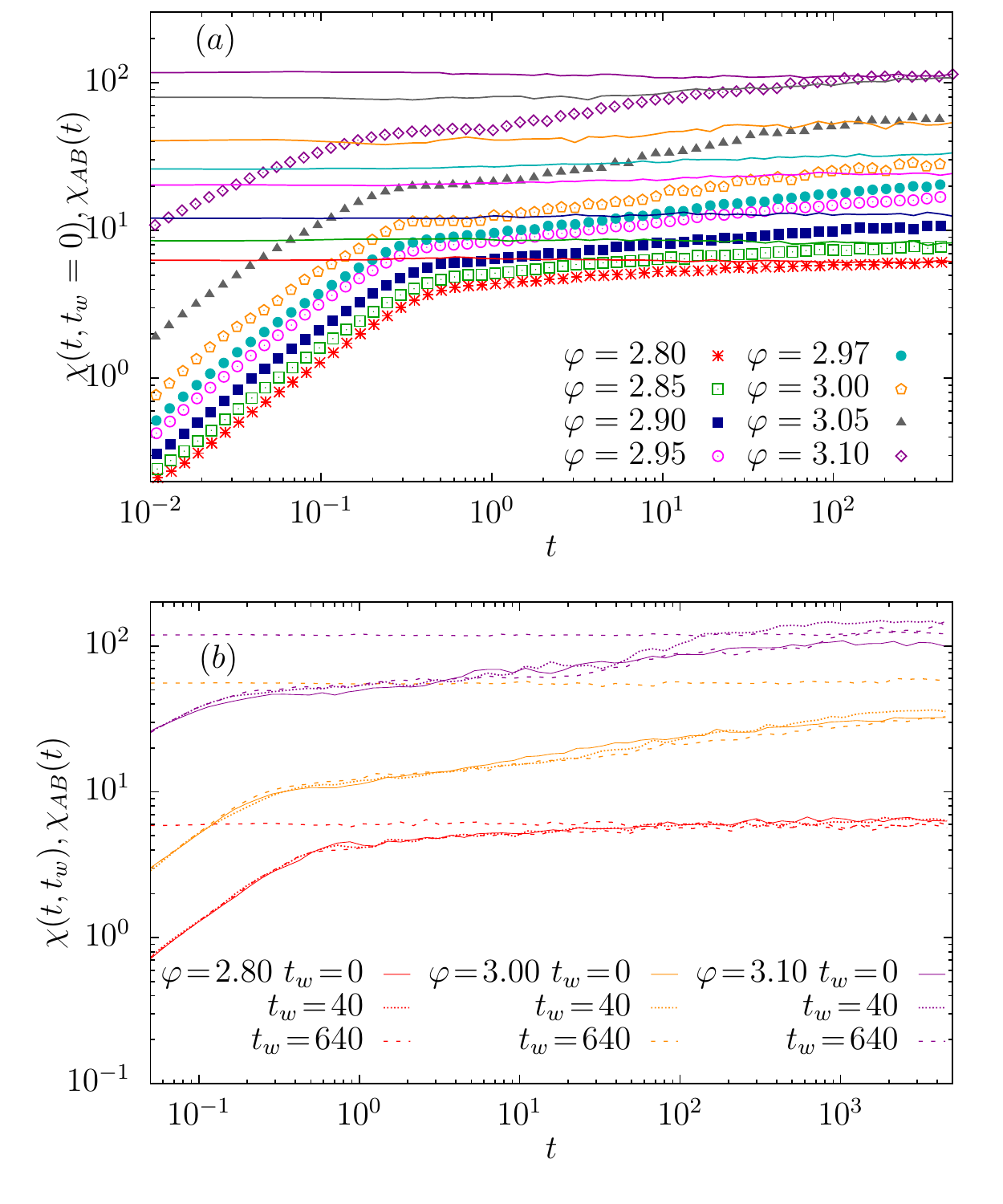}
\caption{(Color online). (a) Time evolution of the caging susceptibility $\chi(t,t_w=0)$ (scatters) and $\chi_{AB}(t)$ (lines). 
(b)~For three selected densities, we report the evolution of $\chi_{AB}(t)$ for much longer times (dotted lines) and the dependence
of $\chi(t,t_w)$ on both $t$ and $t_w$ (see legend).
}
\label{fig:chi-t}
\end{figure}

The dynamical behavior of the caging susceptibility is qualitatively similar {that} of the MSD (Fig.~\ref{fig:chi-t}). 
As for the cage order parameters, {in the SF-RS phase (for $\ph< \phiG$)} the two susceptibilities become identical in the long time limit, i.e., $\chi(t \to \infty,t_w) = \chi_{AB}(t \to \infty)$. By contrast, in the {SF-fullRSB} phase (for $\ph>\phiG$), on a timescale $t  \lesssim \t_{\rm meta}(t_w)$, we generally observe that $\chi(t,t_w)<\chi_{AB}(t)$. Note that the magnitude of the susceptibility increases by more than a decade in the density range considered here, which is a clear signature of the Gardner transition. A detailed analysis of this increase is discussed in Sec.~\ref{sec:susceptibility}.

\subsubsection{Computation of timescales}

Based on the qualitative picture of the dynamics presented above, we now attempt
a more quantitative analysis based
on the {classic work of Ogielski on spin glasses}~\cite{Ogielski1985} 
{and its recent extension} to the study of the dynamical transition in the $d=3$ Edwards-Anderson model under 
an external field~\cite{janus:campodyn}.
The idea consists in obtaining a relaxation timescale $\tau(t_w)$  from the decay of $\d\Delta(t,t_w)$ (Fig.~\ref{fig:delta-t}c).
Note, however, that this scheme is only well defined in the SF-RS phase, where equilibrium can be reached and
Eq.~\eqref{eq:longdD} holds. In the SF-fullRSB phase, $\d\Delta(t,t_w)$ does not decay to zero and evolves continuously over
a broad range of time scales. Extracting a single timescale is {then not so} straightforward. 
We will focus in the following on the data for $t_w=0$, {for which} we have more statistics. {For $\f \lesssim \phiG$}, these data
are {also} representative of all $t_w$.

In order to facilitate the numerical analysis, we make use of analytical results. According to the general theory of
critical glassy dynamics developed in Refs.~\cite{SZ82,CFLPRR12,PR12}, upon approaching the Gardner point in (restricted) equilibrium
from the SF-RS
phase, hence when $\d\D(t,t_w) = \d\D(t)$ does not depend on $t_w$, one has:
\beq
\d\D(t) \sim \d\f \FF( t /\t_\b ) \ , \hskip20pt \t_\b \sim \d\f^{-\g} \ ,
\eeq
where $\d\f = |\f - \phiG|$ and the function $\FF(x)$ is such that $\FF(x\ll 1) \sim x^{-a}$ while $\FF(x\gg 1)$ decays exponentially.
Here the exponents $a$ and $\g = 1/a$ are related to the so-called MCT exponent parameter $\l$ by the relation
\beq
\l = \frac{\G(1-a)^2}{\G(1-2a)} \ .
\eeq
The parameter $\l$ can be computed analytically within the replica method by analizing the cubic terms of the replica 
action~\cite{CFLPRR12,PR12,KPUZ13}. %We performed this computation, whose 
{The results of this computation} are reported in Table~\ref{tab:ana}
(details {are} given in Refs.~\cite{Rainone2666}).

\begin{figure}[t]
\includegraphics [width = \columnwidth] {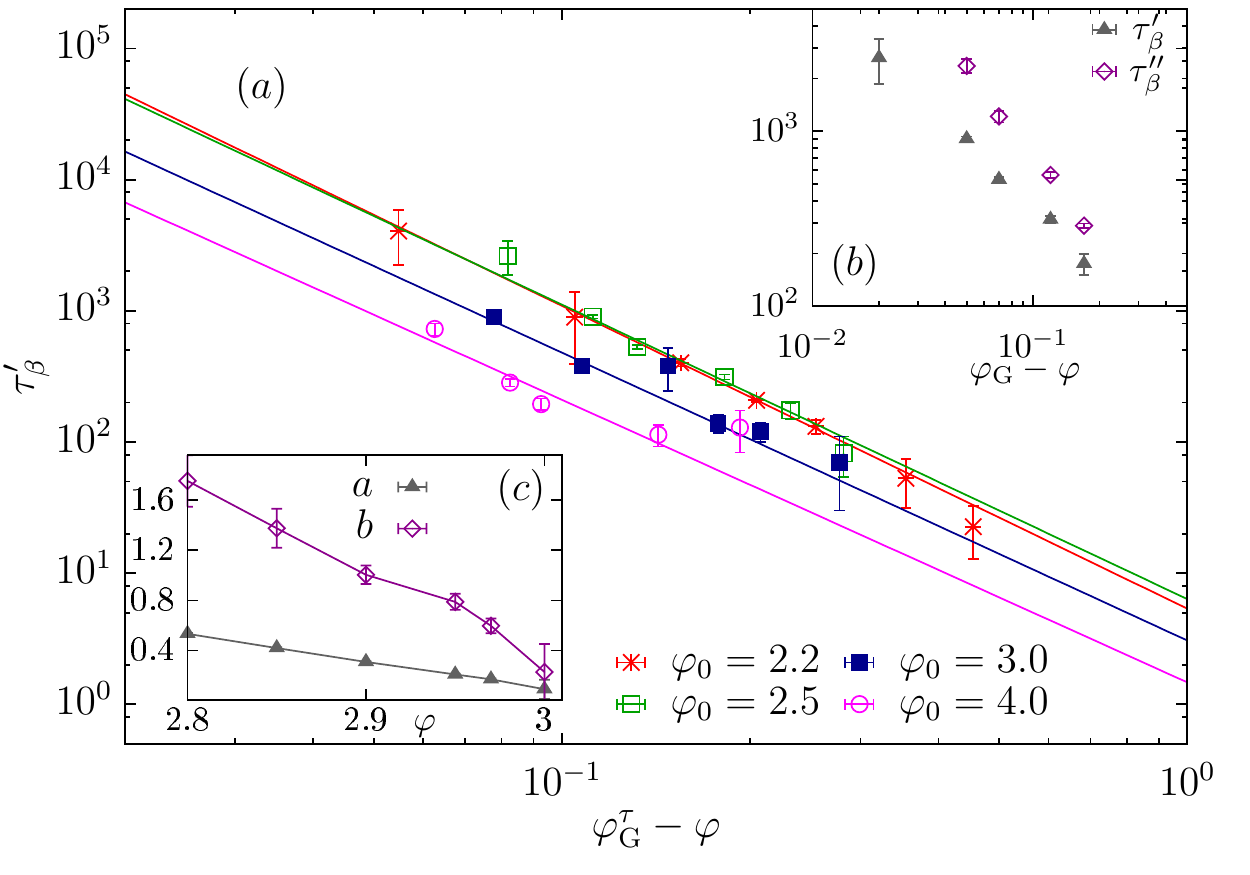} 
\caption{(Color online). (a) Growth of $\tau'_\b$ with $\varphi_\mathrm{G}^\tau-\varphi$ for $\varphi_\mathrm{G}^\tau$ from Table~\ref{tab:phiG}. 
 (b)~The two estimates of $\tau_\b$,  $\tau'_\b$ and $\tau''_\b$, as a function of $\varphi_\mathrm{G}^\tau-\varphi$ for $\phiin=2.5$.
 (c)~Evolution of $a$ and $b$ with $\varphi$ for $\phiin=2.5$. 
}
\label{fig:timescale}
\end{figure}

In order to estimate $\tau_\b$, we fit the results for $\d\D(t,t_w=0)$ using an empirical form $\FF(x) \propto x^{-a}e^{-x^b}$ that has been
used for spin glasses~\cite{Ogielski1985,janus:campodyn}, {which gives} %It results in
\begin{equation}
\d\Delta(t, t_w=0)= 
  c\  \frac{\exp{[-(t/\tau'_\b)^b]}}{t^{a}} \ ,
\label{eq:fit_tau}
\end{equation}
where the parameters $a$, $b$ and $\tau'_\b$ depend on $\ph$, $\phiin$, {and}
$\tau'_\b$ offers a first estimate of $\tau_\b$. 
Note that we fit the exponent $a$ instead of using the analytical result, because the critical regime {over which}
the exponent coincides with $a$ is narrow and away from $\phiG$ the effective exponent is quite different (see Ref.~\cite[Fig.~12]{Ogielski1985}). 
With this choice
all the fits are very good, as reflected by the Pearson 
$\chi_{\rm P}^2$ per degree of freedom (d.o.f.)~\footnote{Recall that 
$\chi_{\rm P}^2=\sum_{i=1}^{N_T}[y_i-f(t_i)]^2/\sigma_i^2$,
where $N_T$ is the numbers of times $t_i$ and $y_i=\d\Delta(t_i,t_w=0)$, $\sigma_i$ is the error of $y_i$, and $f$ is the fitting
function, Eq.~\eqref{eq:fit_tau}} being much less than 1 
(examples of the quality of the fit are given in Fig.~\ref{fig:delta-t}c).
As $\ph$ approaches $\phiG$ for a given $\phiin$, $\tau_\b$ is expected to diverge {as} %according to the power-law form
\begin{equation}\label{fit:tau}
\tau_\b \sim |\ph - \phiG^\tau|^{-\gamma}.
\end{equation}
%Here, 
{In order} to obtain a more constrained value of $\phiG^\t$, we fix the exponent $\gamma$ to its analytic value given in Table~\ref{tab:ana}
and only fit $\phiG^\t$ and the prefactor. 
This time the fits are not excellent, with values of $\chi_{\rm P}^2/\text{d.o.f.}$ of the order of 1 or larger 
(see Fig.~\ref{fig:timescale}). The results for $\phiG^\t$ are reported in Table~\ref{tab:phiG}.

\begin{table}[t]
\begin{tabular}{c@{\hspace{1cm}} c@{\hspace{1cm}}c @{\hspace{1cm}}c}
\hline
$2^d \varphi_0 /d$ & $2^d\phiG/d$ & $\l$ & $\g = 1/a$\\
\hline
4.8 & 4.8 & 0.7027 & 3.069 \\
4.9 & 5.34 & 0.5607 & 2.651\\
5 & 5.64 & 0.5091 & 2.547\\
5.25 & 6.18 & 0.4378 & 2.427\\
5.5 & 6.61 & 0.3938 & 2.363\\
6 & 7.33 & 0.3398 & 2.295\\
6.667 & 8.16 & 0.2957 & 2.245\\
7 & 8.54 & 0.2801 & 2.228 \\
8 & 9.63 & 0.2469 & 2.194 \\
10.667 & 12.36 & 0.2042 & 2.154 \\
\hline
\end{tabular}
\caption{Analytical results (in the limit $d\to\io$) for $\phiG$ and the exponents $\l$, $\g$ and
$a$ for several $\phiin$~\label{table:lambdaMCT}.}
\label{tab:ana}
\end{table}

%\begin{table}[b]
%\caption{Fit parameters of Eq.~(\ref{fit:tau}) for different $\phiin$.
%}
%\label{tab:fittau}
%\begin{tabular}{ccc}
%\hline
% $\phiin$&$\phiG^{\tau}$&$\chi_{\rm P}^2/\text{d.o.f.}$,\\\hline
% 2.2&2.755(5)&0.9/5,\\
% 2.5&3.082(13)&18.9/4,\\
% 3.0&3.576(11)&9/4,\\ 
% 4.0&4.59(2)&19.7/3.\\ 
% \hline
%\end{tabular}
%\end{table} 

\begin{figure}[t]
\includegraphics [width = \columnwidth] {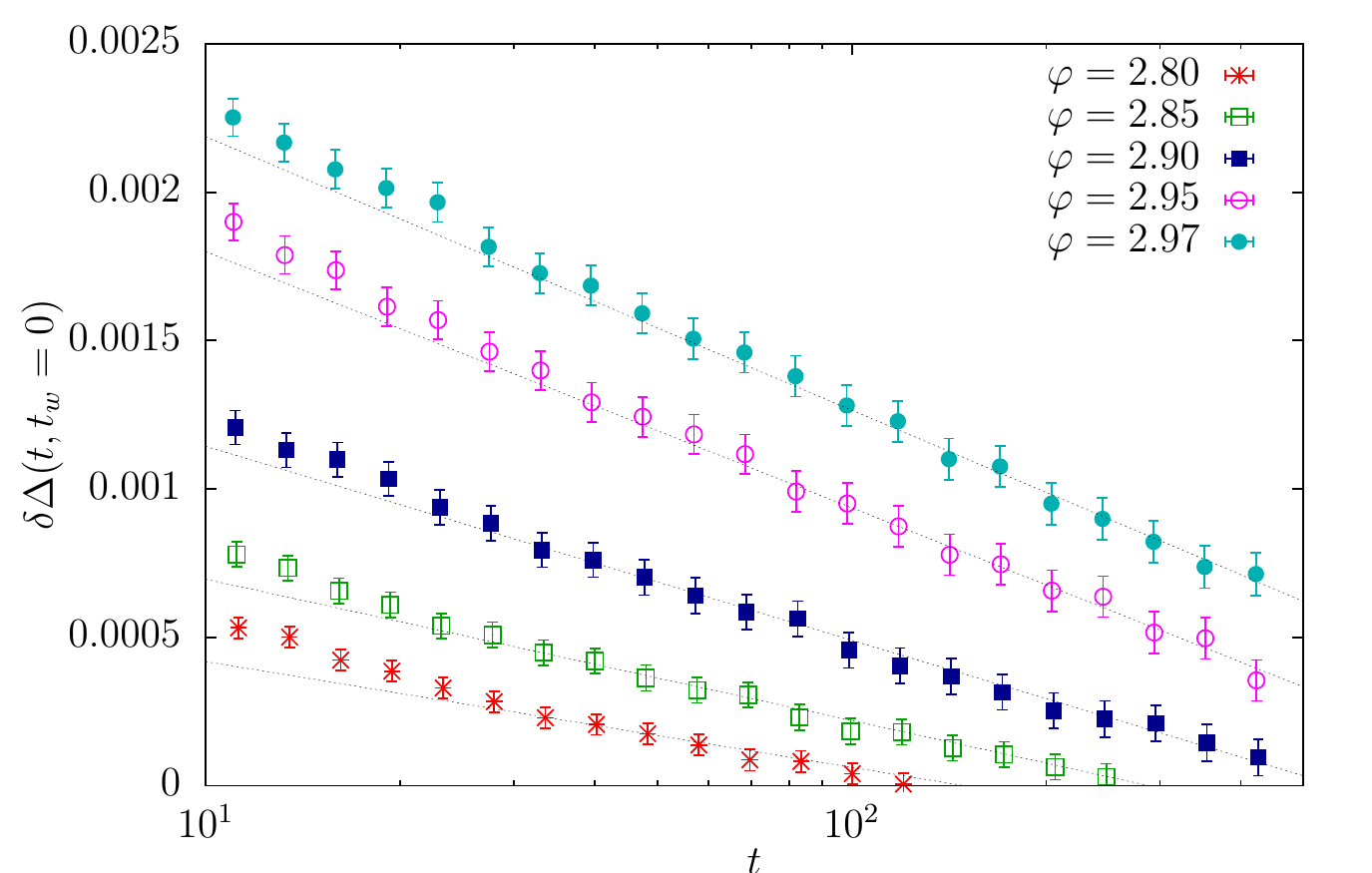} 
\caption{(Color online). 
%{\color{red} Make this figure with a regular log scale, not log10(t)} 
Logarithmic decay of $\d\Delta(t,t_w=0)$ with time for $\phiin=2.5$. The results are fitted to Eq.~\eqref{eq:fit_tau2} (dashed lines).}
\label{fig:timescale-alt}
\end{figure}

An alternative estimate of $\tau_\b$ can be obtained from the logarithmic 
scaling of $\d\Delta(t,t_w=0)$ at long times (see
Fig.~\ref{fig:timescale-alt}). For
$\ph<\phiG$, the fitting form
\begin{equation}\label{eq:fit_tau2}
\d \Delta(t,t_w=0)= k \left[1-\frac{\log (t)}{\log (\tau''_\b)}\right]
\end{equation}
with a density-dependent constant $k$ gives $\tau''_\b$~\cite{janus:campodyn}. 
Comparing $\tau'_\b$ and $\tau''_\b$ suggests that the divergence of the two
timescales is compatible with a same $\phiG$ and a very
similar power-law exponent (see
Fig.~\ref{fig:timescale}b). The insensibility of the estimator of $\tau_\b$ to its
precise definition adds support to our claim that the observed
divergence is due to a true thermodynamic transition. 
Also, we repeated the analysis for $t_w > 0$ with very similar results.

To conclude the discussion, note that the results for $\DAB(t)$ and $\Delta(t,t_w)$ at low
densities, i.e., $\phid\leq\phiin \lesssim 2.2$, may be affected by hopping (Sec.~\ref{sec:finiteD} and~\cite{Charbonneau2014}). In these systems the timescale for leaving a metabasin (albeit only through local hopping processes) is comparable to $\tau_\b$ even near $\phiG$. The estimate of $\phiG$ in this regime is therefore subject to a larger error, which explains the bigger difference between $\varphi_\mathrm{G}^{\tau}$
and other $\phiG$ estimates (Table~\ref{tab:phiG}). For the limit case $\phiin = 1.8$, we do not even attempt to fit the data because
no clear power-law regime can be distinguished. By contrast, for $\phiin \ge 2.5$, hopping is negligible on the timescales achieved numerically.

\subsection{Static functions}
\label{sec:static}

As we have shown in Sec.~\ref{sec:time}, at the Gardner transition two-clones observables such 
as $\DAB(t)$ become different from the long-time limit of dynamic observables such as $\D(t,t_w)$.
In this subsection, we thus estimate the location of the Gardner transition using an approach based on the study of
two-clone static observables. %We call them 
{These observables are} static because,
as we showed in Sec.~\ref{sec:time}, they are time independent for
$\varphi < \phiG$. {We can thus arbitrarily choose any $t$} to compute them. 
Here, we choose $t_{\rm s}=0.2V^{1/3} \sim 2$, such that
$\tau_0 < t_{\rm s} \ll \tau_{\rm meta}$, {which we abbreviate below as}
$\Delta\equiv\Delta(t_{\rm s},t_w=0)$ and
$\DAB\equiv\DAB(t_{\rm s})$. {We use a similar} notation for all other observables 
unless otherwise specified.

\subsubsection{Average mean square displacement}
\label{sec:avmsd}

\begin{figure}[t]
\includegraphics [width = \columnwidth]{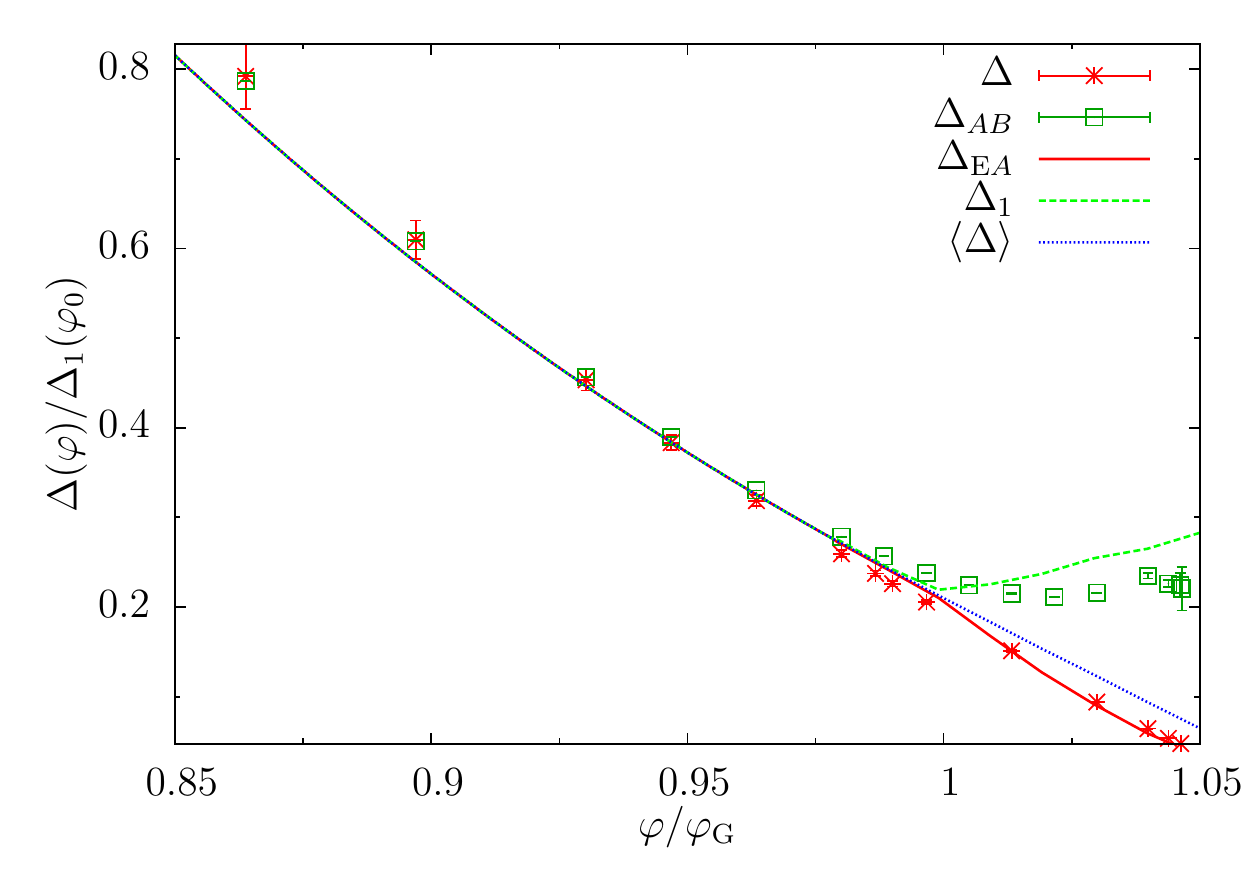}
\caption{(Color online). Density evolution of the averages of $\Delta$ and $\Delta_{AB}$ for $\phiin=2.50$. 
Numerical results (points) are compared with theoretical predictions for $\D_1$, $\D_{\rm EA}$ and $\langle \D \rangle_{\rm SF}$ (lines). 
Both datasets are scaled to reference values (see Sec.~\ref{sec:avmsd} for details) to obtain a better agreement between theory and simulation.
}
\label{fig:delta_phi}
\end{figure}

%We begin the discussion by looking at 
{We consider the averages of $\D$ and $\DAB$ and compare them with the theoretical SF results (see Fig.~\ref{fig:delta_phi}
for $\phiin=2.5$)}.
In order to take into account the corrections discussed in Sec.~\ref{sec:finiteD}, both the numerical and the theoretical
datasets have been rescaled. For the vertical axis, we rescale {the results to $\D_1(\phiin)$, which} is known both
in theory and in simulation, {such} that both datasets are equal to 1 for $\f = \phiin$. 
For the horizontal axis, we rescale $\f$ to $\phiG$ %for the theoretical curve the 
{with the theoretical $\phiG$ for the SF-RS curve} (Table~\ref{table:lambdaMCT}) {and a fit factor
$\phiG^{\rm T}$ for the numerical results} (Table~\ref{tab:phiG}). % to obtain the best superposition of the datasets.
%The resulting 
Figure~\ref{fig:delta_phi}
clearly shows the difference between the two expected regimes (Sec.~\ref{sec:gardner}):
for $\f < \phiG$, we {obtain} $\D \sim \DAB \sim \D_1$, while for $\f>\phiG$, we {obtain} $\D \sim \DEA$ and $\DAB \sim \D_1$,
and hence $\DAB > \D$.
%Note that while 
{Although} at short times we observe $\DAB(t) \approx \D_1$, 
on much longer timescales we expect $\DAB(t)$ to evolve slowly towards its equilibrium value $\DAB(t) = \langle \D \rangle_{\rm SF}$,
which is however only approached for $t$ that diverge with $N$.

\subsubsection{Probability distribution functions}

{We next consider the probability distribution function (pdf)}
%Next, we study the pdf 
of the cage order parameters by computing $\Delta$ and $\DAB$ for each
sample and constructing the histogram over samples.
Note that because
$\Delta$ is the mean square displacement restricted to a single {sub-basin and the waiting time is fairly short}, its distribution
represents $P_{\rm SF}(\Delta)$ in
the SF-RS phase, {and only} %while it represents 
the peak around $\Delta_{\rm EA}$ in the SF-fullRSB
phase (see Fig~\ref{fig:fractal}).

\begin{figure}[t]
\begin{center}
\includegraphics [width = \columnwidth] {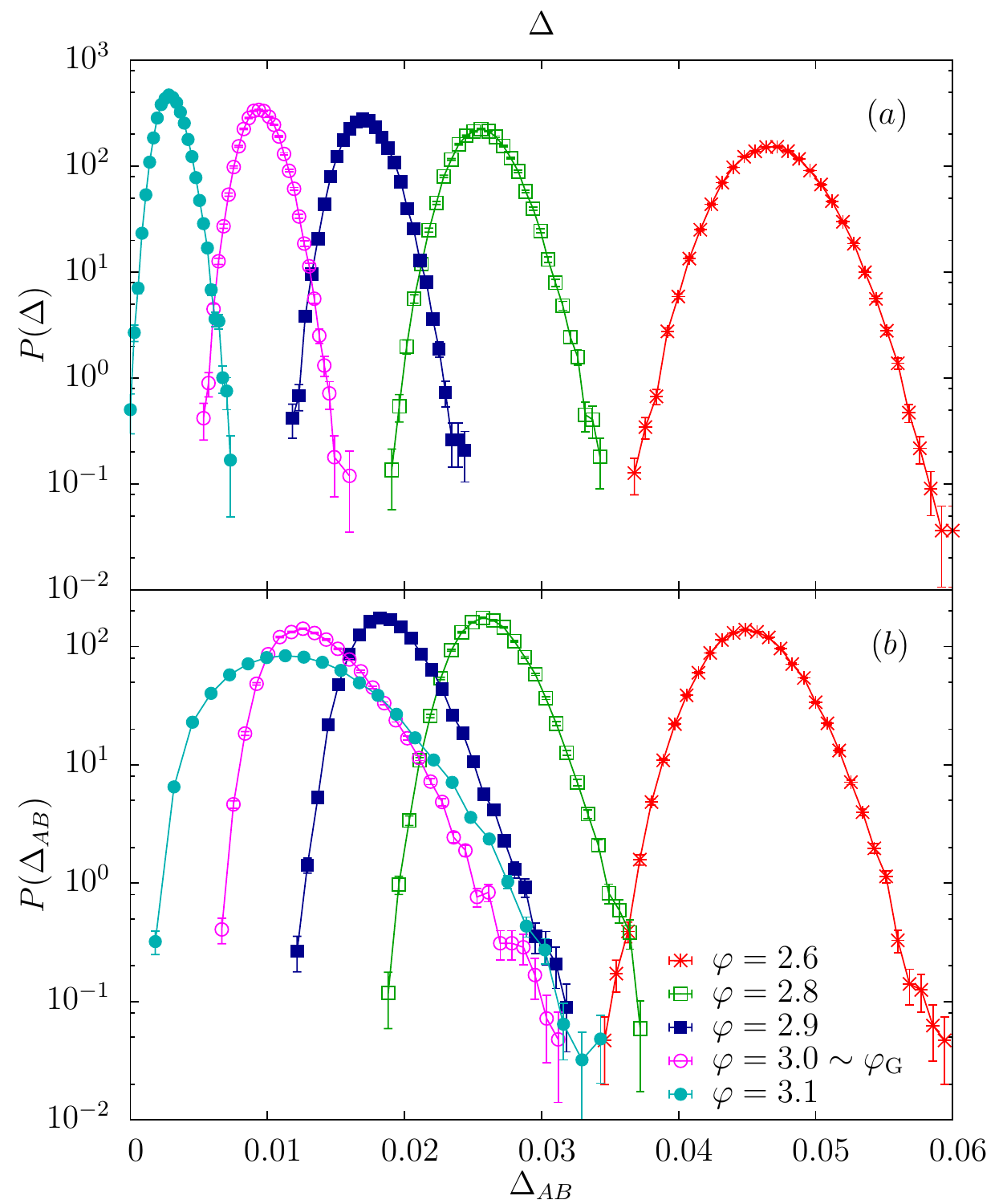}
\end{center}
\caption{(Color online). (a) $P(\Delta)$  and (b) $P(\DAB)$ at different $\ph$ for $\phiin=2.5$.
} \label{fig:pdf}
\end{figure}

Figure~\ref{fig:pdf} shows $P(\Delta)$ and $P(\DAB)$ for $\phiin=2.5$ calculated
from
$N_{\mathrm{s}}=40,000$ -- 75,000 samples. The shape of $P(\Delta)$ is Gaussian-like 
at all $\ph$, and {its} mean value monotonically 
decreases with increasing $\ph$. The shape of $P(\DAB)$, however, changes
considerably over that same regime. For $\f < \phiG$, it is Gaussian
and analogous to that of $P(\Delta)$, but 
near $\phiG$ it develops an exponential tail
akin to a Gumbel distribution. If
$\varphi$ is further increased, $P(\DAB)$ then becomes broader, which is consistent with the 
theoretical expectation
 (see Fig.~\ref{fig:fractal}).

\begin{figure}[t]
\begin{center}
\includegraphics [width = \columnwidth] {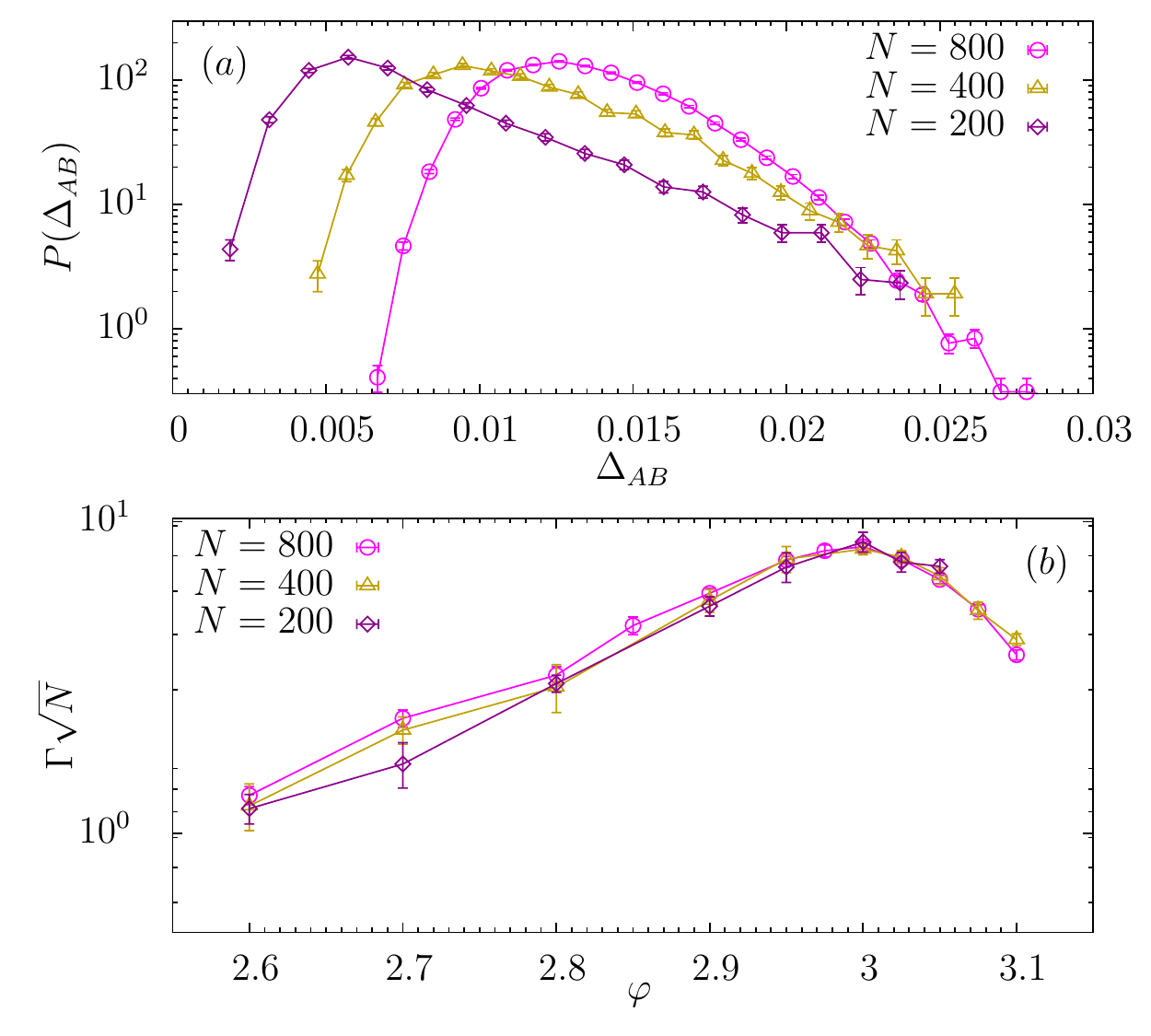}
\end{center}
 \caption{(Color online). (a) Finite-size behavior of $P(\DAB)$ at $\ph = 3.0 \approx \phiG$ and (b) density evolution of the rescaled skewness $\Gamma$ for three different $N$ and for $\phiin=2.5$.
 }\label{fig:SKpdf}
\end{figure}

The development of an exponential tail at the critical point has been
observed and studied for spin glasses in a
field~\cite{Pa12,Janus14}. The effect is thought to be due to disorder. 
Whereas the results for most samples fall within Gaussian
fluctuations around a given mean value, a few rare samples have 
much larger $\DAB$ than the mean, giving an 
exponential tail to the distribution. The smaller the system, the stronger the effect (Fig.~\ref{fig:SKpdf}). 
These rare fluctuations are hypothesized to originate from the sample-to-sample fluctuations of the critical
point~\cite{Pa12}, which then translates into significant sample-to-sample
fluctuations of some of the measured observables. 
We come back to
this point in Sec.~\ref{sec:sample}.

The connection between the changing shape of the distribution and criticality suggests that we can determine the critical transition from $P(\DAB)$ alone. We propose below two alternative procedures for detecting the Gardner transition using standard moments of the distribution. 

\subsubsection{Caging susceptibility}
\label{sec:susceptibility}

We first define a caging susceptibility from the normalized
variance of $P(\DAB)$ 
\begin{equation}\label{eq:chi}
\chi=N\frac{\av{\DAB^2}-{\av{\DAB}^2}}{\av{\DAB}^2},
\end{equation}
where the denominator corrects for the fact that $\av{\DAB}$ changes with $\ph$. As in the vicinity of any critical point, the susceptibility is expected to diverge as
\begin{equation}\label{eq:divergence}
\chi\propto (\phiG^\chi-\varphi)^{-1},
\end{equation}
where the critical exponent 1 is due to the fact that the MK model is mean-field in nature.

\begin{figure}[t]
\begin{center}
\includegraphics [width = \columnwidth] {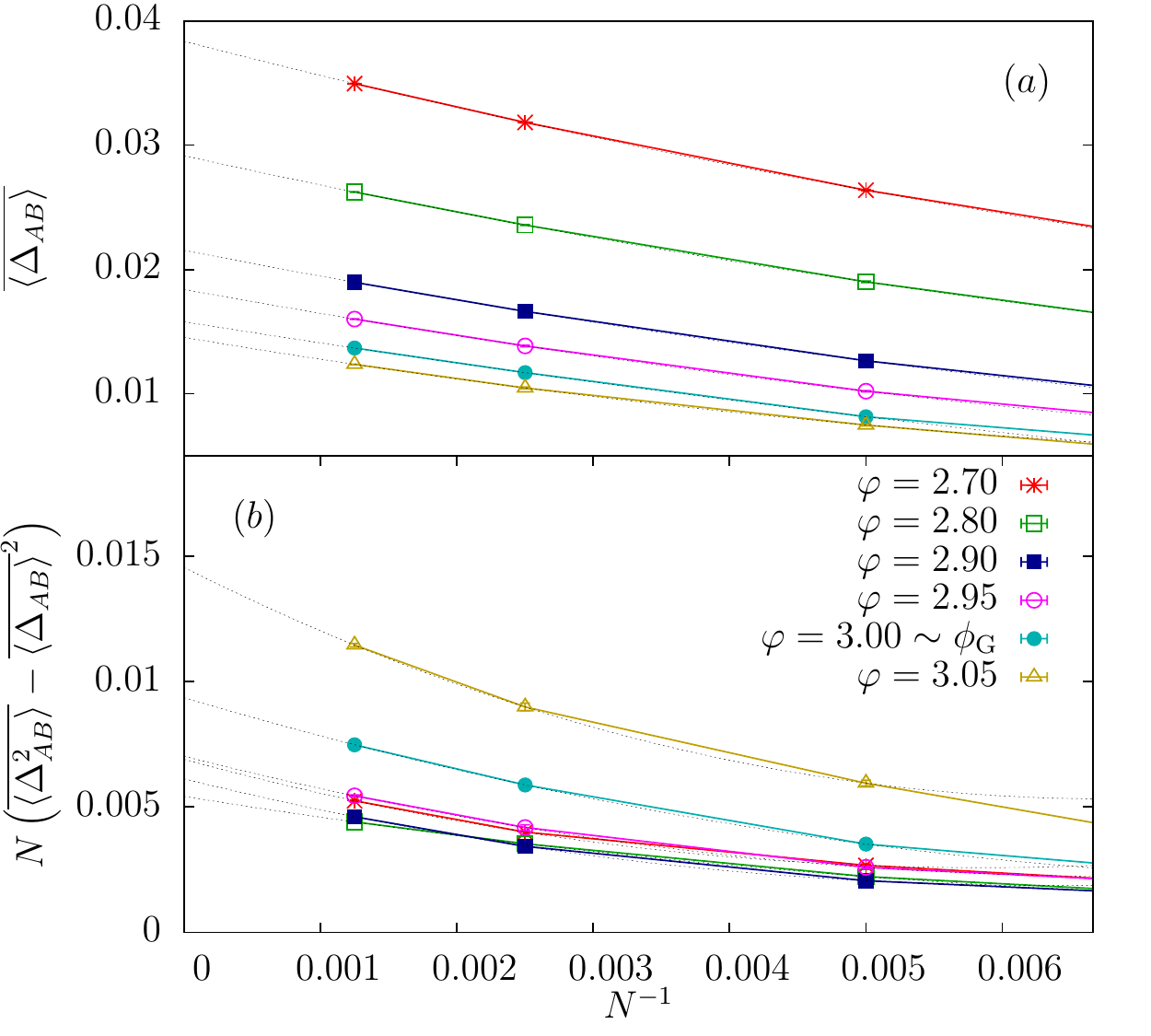}
\end{center}
 \caption{(Color online). Finite-size behavior of the two terms of the definition of the
   susceptibility in Eq.~\eqref{eq:chi}: (a) $\av{\DAB}$, and (b)
   $N(\av{\DAB^2}-{\av{\DAB}^2})$, for $\phiin=2.50$. The dashed lines are
   fits of a second-order polynomial in
   $1/N$ to the data.  The $1/N\to0$ extrapolation of these two fits have been used to extract the
   ${N\to\infty}$ limit of $\chi$, given in Fig.~\ref{fig:chiSK} by a
   dark solid line.}\label{fig:finite-size-chi}.
\end{figure}

The definition of $\chi$ involves taking the quotient of two quantities that
  both suffer from strong finite-size corrections. In order to control for this effect
  we study the behavior of both terms as function of $1/N$ (Fig.~\ref{fig:finite-size-chi}). The denominator, $\av{\DAB}$, behaves as a regular 
  series in $1/N$, decreases smoothly with $\ph$, and eventually saturates above the
  Gardner point. The numerator, 
  $N(\av{\DAB^2}-{\av{\DAB}^2})$ has, however, a more complex behavior. While it follows a
  nearly $1/N$ behavior for $\ph<\phiG$ with a small dependence on $\ph$, it
  grows significantly faster both with $N$ and $\ph$ for $\ph>\phiG$. 
  We attempt to
  extract the value of both quantities at the thermodynamic limit using a second-order
  polynomial fit in $1/N$, even {though} for $\ph > \phiG$ the estimate of $N(\av{\DAB^2}-{\av{\DAB}^2})$
  obtained in this way is not reliable and {larger systems would be needed} to obtain a better extrapolation.

\begin{figure}[t]
\begin{center}
\includegraphics [width = \columnwidth] {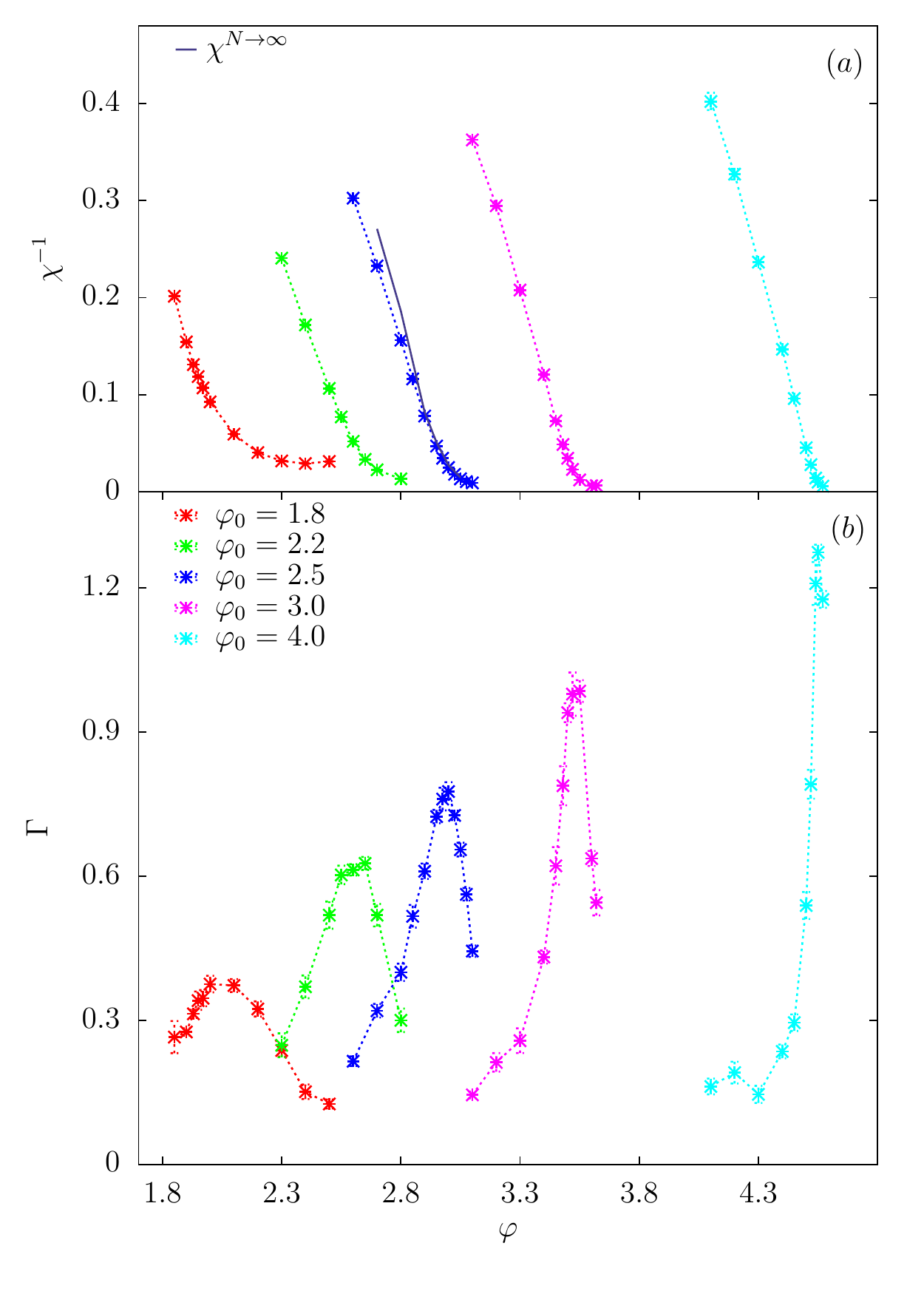}
\end{center}
 \caption{(Color online). (a) Inverse caging susceptibility, $\chi^{-1}$, and (b)
   skewness, $\Gamma$, as functions of $\varphi$ for different $\phiin$. Upon
   approaching $\phiG$, $\chi$ apparently diverges as
   $|\varphi-\varphi_\mathrm{G}^\chi|^{-1}$, except for $\phiin=1.8$. Note
   that the divergence of $\chi$ coincides with the maximum of $\Gamma$. 
        The    $N\to\infty$ extrapolation of the curves in 
     Fig.~\ref{fig:finite-size-chi} were used to obtain the dotted line.}
     \label{fig:chiSK}
\end{figure}

The results for the susceptibility for $N=800$ are reported in 
Fig.~\ref{fig:chiSK}a for different values of $\phiin$ (and thus $\phiG$).
  For $\phiin=2.50$ we also include $\chi^{N\to\infty}$ obtained using the
  extrapolations of Fig.~\ref{fig:finite-size-chi}. The comparison suggests that although the finite-size
  effects in both $\av{\DAB}$ and $N(\av{\DAB^2}-{\av{\DAB}^2})$ are still very strong for this system size, 
  the determination of $\chi$ is fairly well controlled, at least for $\f < \phiG$.

{The numerical data in Fig.~\ref{fig:chiSK}a suggest} a roughly linear behavior of $\chi^{-1}$ for all $\phiin$ except 
$\phiin=1.8$, where the spacing between $\phiin$ and $\phiG$ is narrowest, and where hopping most likely obfuscates the critical regime (see Fig.~\ref{fig:phase_diagram}). 
%The linear behavior is better observed the higher $\phiin$ is. 
For the other $\phiin$, we {estimate} the Gardner transition by fitting Eq.~\eqref{eq:divergence} in the $\ph<\phiG$ region (Table~\ref{tab:phiG} and Fig.~\ref{fig:phase_diagram}).

\begin{figure}[t]
\begin{center}
\includegraphics [width = \columnwidth] {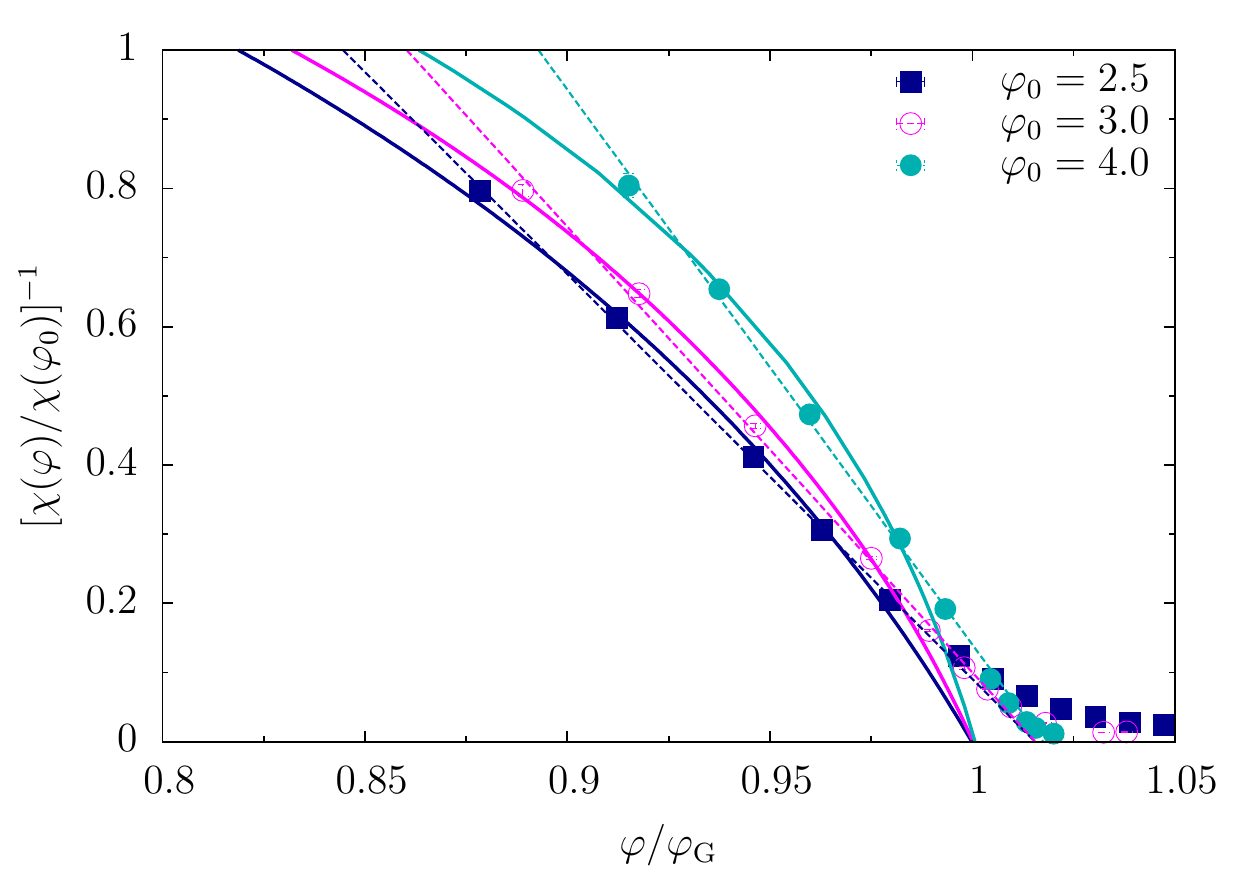}
\end{center}
 \caption{(Color online). 
 The numerical susceptibilities (scatters) are compared with theoretical predictions (lines). 
 Both datasets are scaled to reference values (see Sec.~\ref{sec:susceptibility} for details) to obtain a better agreement between theory and simulation.
 We also report the linear fits to the numerical data used to determine $\phiG^\c$ (dashed lines). 
This linear fit overestimates the transition point by
     roughly~2\%. 
     }\label{fig:chiSK_compare}
\end{figure}

In Fig.~\ref{fig:chiSK_compare} we compare the numerical {results} with the theoretical predictions.
{Note, however, that the theoretical curves} do not correspond to the full $\chi$ {value},
but only {include} the leading divergent term, namely the inverse of the replicon eigenvalue
(see~\cite{Charbonneau2014B,Rainone2666} for details).
{As discussed for Fig.~\ref{fig:delta_phi}, %in order to obtain a better matching and take into account the corrections discussed in  and discussed in Sec.~\ref{sec:finiteD}, 
both the theoretical and numerical
datasets are {also} rescaled. The vertical axis is normalized to {1 at} $\f=\phiin$, while the horizontal $\f$
axis is scaled to $\phiG$, using the theoretical 
values in Table~\ref{table:lambdaMCT} and the fit values
$\phiG^{\rm T}$ for the numerics (Table~\ref{tab:phiG})}. % to obtain the best superposition of the datasets.
{Interestingly,} the theoretical curves are not linear {over} the whole density regime. {They instead} bend towards $\phiG$, 
before {vanishing linearly, but only in a fairly} small region around $\phiG$. {The linear scaling observed in the simulation data is thus likely} %display
%an apparent linear
%behavior over a much larger scale, which is however 
due to finite-size effects. 
%As a result, 
The numerical estimates of $\phiG^\c$ obtained {from the}  linear fit of the whole set of data for $\f < \phiG$ 
{thus} slightly
overestimate the true transition point. {Yet the effect is quite small -- approximately $2\%$, based on Fig.~\ref{fig:chiSK_compare}}.
For $\f \sim \phiG$ and above, we {observe} that finite-size effects dominate the numerical determination
of $\chi$, which remains finite instead of diverging. % as predicted by the theory.

\subsubsection{Caging skewness}

Near the Gardner transition, large sample-to-sample fluctuations give rise to a strong exponential tail in  $P(\DAB)$. This effect can be quantified by the skewness of the distribution
\begin{equation}\label{eq:skewness}
\Gamma=\frac{\av{\left(\DAB-{\av{\DAB}}\right)^3}}{\av{\left(\DAB-\av{\DAB}\right)^2}^{3/2}}\ .
\end{equation}
Recall that the skewness is a measure of a distribution's asymmetry
and that a Gaussian distribution would have $\Gamma=0$.  Sample-to-sample fluctuations are expected to be maximal at the critical point (see Sec.~\ref{sec:sample}), which provides an estimate of the Gardner transition, $\phiG^{\Gamma}$ (see Fig.~\ref{fig:chiSK}b and Table~\ref{tab:phiG}).
For all $\phiin$, we see that
$\phiG^{\Gamma}$ is very close to the fitted divergence of the susceptibility, $\phiG^\chi$. 
Finite-size analysis further shows
that the rescaled skewness, $\Gamma \sqrt{N}$, collapses the data for different $N$ 
(Fig.~\ref{fig:SKpdf}). The peak
of $\Gamma$ is thus expected to persist all the
way to the thermodynamic limit, consistently with comparable
observations in spin-glass models~\cite{Pa12}.

\subsection{Sample-to-sample fluctuations}
\label{sec:sample}

\begin{figure}[t]
\begin{center}
\includegraphics [width = 0.8\columnwidth] {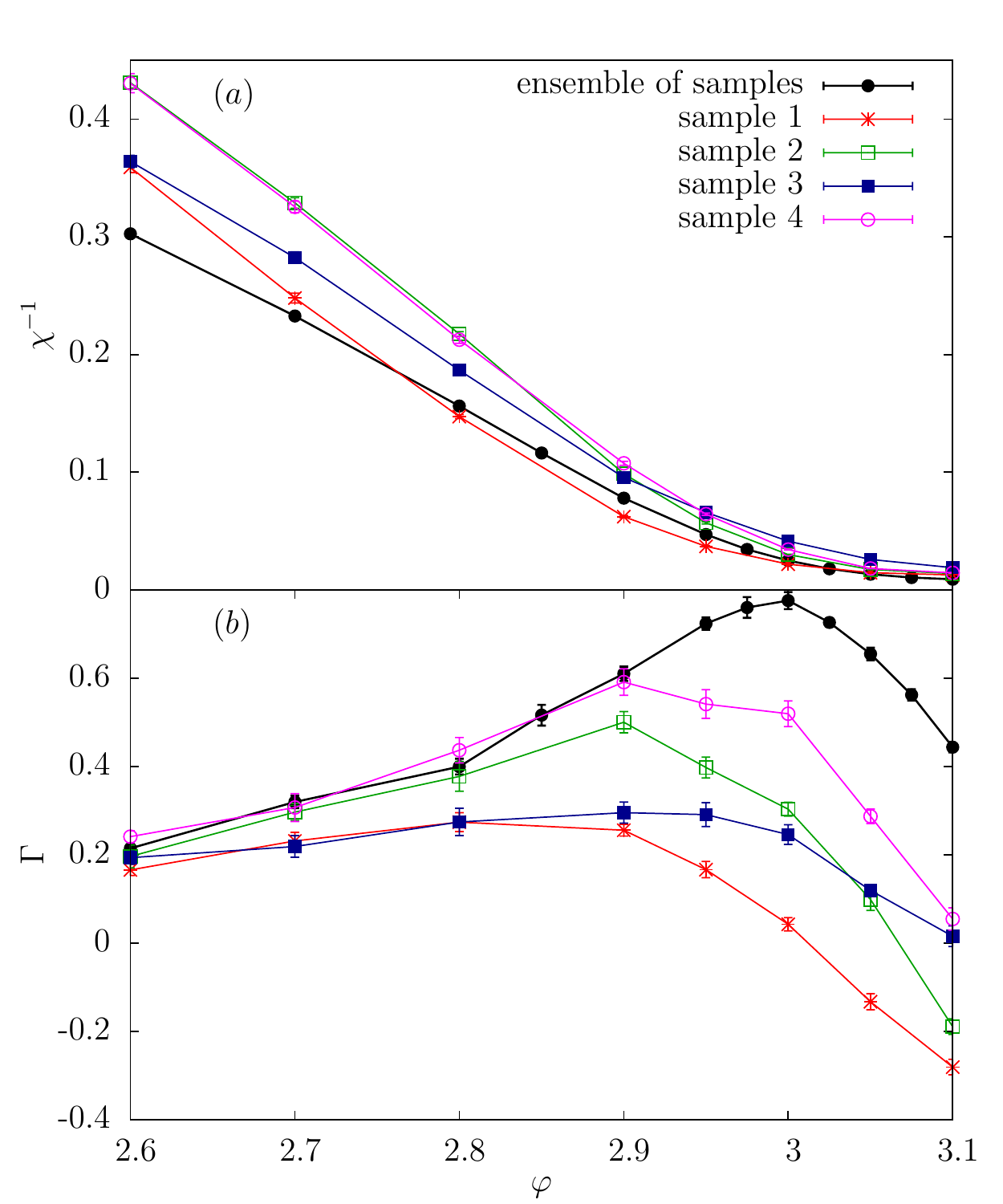}
\end{center}
\caption{(Color online). (a) The susceptibility $\chi$ and (b) the skewness $\Gamma$ averaged over the ensemble of clones are compared with those of four individual samples, for $\phiin=2.50$.
}\label{fig:sample_dependence}
\end{figure}

We have assumed above that the abnormal behavior of  $\Gamma$ around the Gardner transition is due to sample-to-sample fluctuations,
and we further motivate this hypothesis here.
Recall that in our notation a sample is defined by a given configuration
$\{\br_i\}$ and a set of random shifts $\{\boldsymbol{\Lambda}_{ij}\}$ (see Sec.~\ref{sec:planting}).
To compute the moments $\chi$ and $\Gamma$ of each sample, 
we perform $N_{\mathrm s}=10,000$ independent clonings for each sample. 
In Fig.~\ref{fig:sample_dependence}, the data for the ensemble of samples (same data as in Fig.~\ref{fig:chiSK})
are compared with those of four individual samples. We note that the density evolution of $\Gamma$ for the individual samples can have a very different behavior from the ensemble one. In particular, a peak around $\phiG$ is generally not seen. 

Sample-to-sample fluctuations have a smaller effect on $\chi$ (Fig.~\ref{fig:sample_dependence}).  While the magnitude and the critical density exhibit some fluctuations, they all display a divergence similar to that of Eq.~\eqref{eq:divergence}. Because each realization of disorder corresponds to different SF-RS metabasins, our results suggest that the metabasins themselves have slightly different properties. In particular, they exhibit different $\phiG$, which is likely the physical origin of the exponential tail of $P(\Delta_{\rm AB})$ and thus of its anomalous skewness~\cite{Pa12}.

\section{Summary of results}
\label{sec:summary}
\begin{figure}[t]
\centerline{\hbox{ \includegraphics [width = \columnwidth] {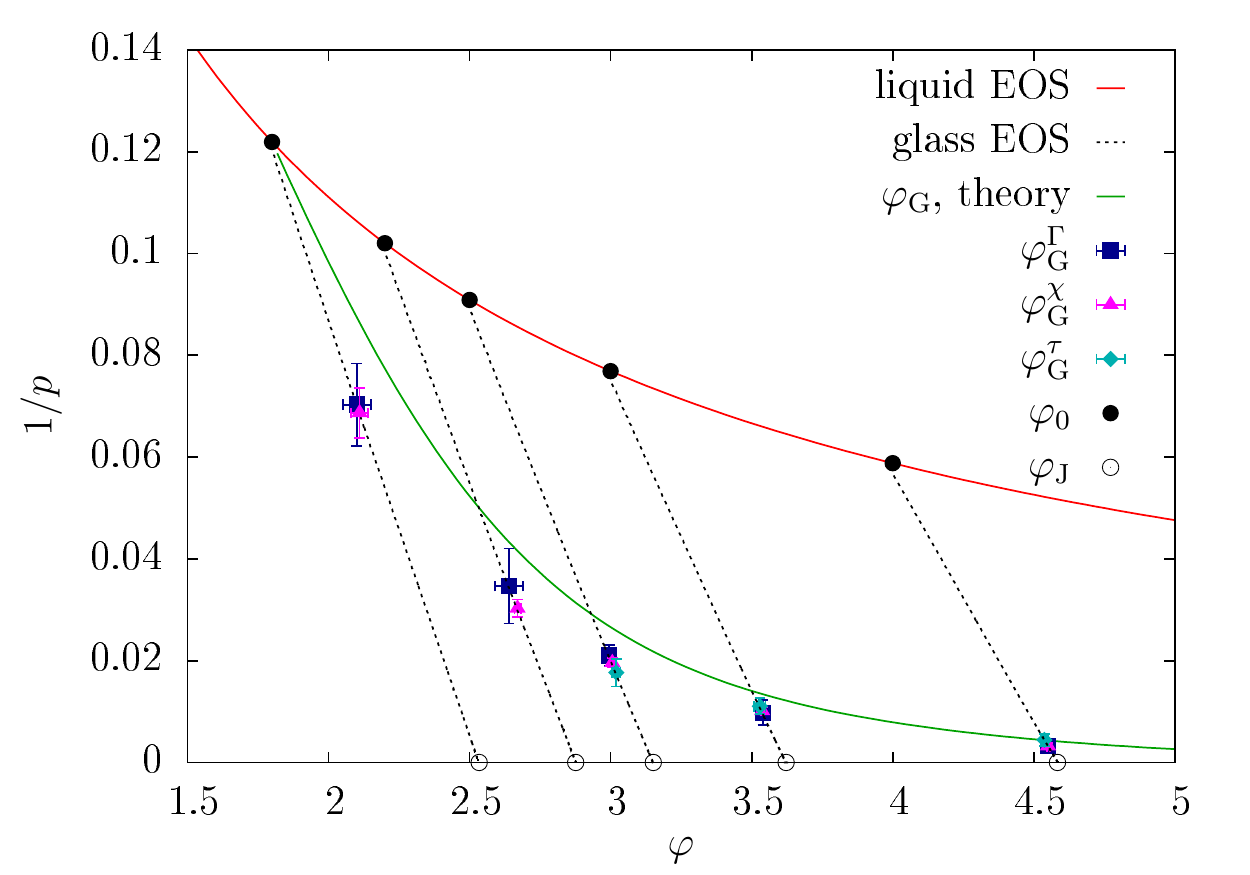}}}
\caption{(Color online). Inverse reduced pressure $1/p$ density $\f$ phase diagram of the MK model in $d = 3$. The liquid EOS follows Eq.~\eqref{eq:liquid_eos}, while a specific state follows a glass EOS from $\phiin$ up to jamming at $\phij$. Numerical estimates of the Gardner transition evolve with $\phiin$ similarly as the theoretical predictions~\cite{Rainone2014}. 
}
 %The theoretical densities, which are obtained in a scaled unit as $\hat{\f} = 2^d \f/d$ for infinite dimensional systems ($d \to \infty$), are rescaled to $d=3$.}
\label{fig:phase_diagram}
\end{figure}

The Gardner transition at $\phiG$ was independently and quantitatively identified from: {\it (i)}~the power-law divergence of the characteristic time $\tau$ in Fig.~\ref{fig:timescale}, {\it (ii)}~the linear vanishing of the inverse susceptibility $\chi^{-1}$ in Fig.~\ref{fig:chiSK}a, 
and {\it (iii)}~the maximum of the skewness $\Gamma$ in Fig.~\ref{fig:chiSK}b. 
Table~\ref{tab:phiG} contains these results and 
Figure~\ref{fig:phase_diagram} {presents} them in a phase diagram. The different estimates of $\phiG$ are generally quantitatively consistent with each other and qualitatively consistent with the mean-field SF results from $d\to\infty$. 

The agreement with the mean-field calculation improves with density, likely because the Gaussian caging approximation used in the theory becomes a better approximation at higher densities~\cite{KPZ12,Charbonneau2014}.
%In the limit of high densities, we find that $\phiG$ is linearly related to $\phiin$ (Fig.~\ref{fig:phG_comparison}), which can be understood from a simple free-volume argument. In fact, it is easy to obtain from Eqs.~(\ref{eq:liquid_eos}) and (\ref{eq:free_volume}) that $\phiG \approx \phij \approx \phiin + \frac{1}{2^{d-1}C}$, where $C$ becomes density independent when $\f \to \infty$. 
A couple of reasons underlie the discrepancy between numerical estimates and theory in the vicinity of $\phid$. First, caging becomes imperfect at such low densities, which allows particles to hop between neighboring cages on  a timescale comparable with the simulation time~\cite{Charbonneau2014}. Hopping thus affects the dynamics of the system ($\Delta(t,t_w)$ and $\DAB(t)$) and also transforms the transitions (at either $\phid$ or $\phiG$) in crossovers. Second, because $\phiG$ is expected to converge to $\phid$ upon approaching the dynamical glass transition  (see  Figs.~\ref{fig:theory} and~\ref{fig:phase_diagram}), the critical regime becomes too small to make any fit to a critical power-law scaling.   
%{\bf \color{red} FZ:Do we really need Fig.~\ref{fig:phG_comparison}? Is it a really useful information that is not already contained in Fig.~\ref{fig:phase_diagram}?}

\begin{table}[t]
\caption{
Results for $\varphi_\mathrm{G}$ and $\phij$
  for  various $\phiin$.
Errors are estimated as follows. 
For $\varphi_\mathrm{G}^{\Gamma}$, the error is the average distance between the susceptibility maximum and the next largest point. 
For $\varphi_\mathrm{G}^{\mathrm{\chi}}$ and $\varphi_\mathrm{G}^{\mathrm{\tau}}$, fitting $\f$ 
intervals are varied to obtain the smallest and largest $\phiG$ for which a fit is possible. The distance between the two values is the error.
  }
\centering
\begin{tabular}{ |c | c | c | c|c| c|}
  \hline
  $\phiin$ & $\varphi_\mathrm{G}^{\Gamma}$ & $\varphi_\mathrm{G}^{\mathrm{\chi}}$&$\varphi_\mathrm{G}^{\mathrm{\tau}}$&$\varphi_\mathrm{G}^{\mathrm{T}}$&$\phij$\\
  \hline\hline
  1.8 & 2.10(5) & 2.11(3) &    --    & -- & 2.534(3)\\
  2.2 & 2.64(5) & 2.670(11) &      2.72(4)& -- & 2.876(4)\\
  2.5 & 2.995(15) &  3.006(6)&3.06(2)&2.96 &3.151(3)\\
  3.0 & 3.54(2) &  3.537(3)&3.554(17)&3.49 &3.622(5)\\
  4.0 & 4.550(12) &  4.551(5) &4.57(2)&4.48&4.584(4)\\
  \hline  
\end{tabular}
\label{tab:phiG}
\end{table}

%\begin{figure}[t]
%\includegraphics [width = \columnwidth]{phiG_comparison.pdf}
%\caption{The Gardner transition $\phiG$ of both simulation and theoretical values are plotted as a function of  the initial density $\phiin$. The asymptotic scaling in the large density limit is fitted to a linear behavior $\phiG \sim \phiin + 0.54$ (line).}
%\label{fig:phG_comparison}
%\end{figure}

In addition to {\it (i)-(iii)},  a {more indirect} way {\it (iv)} to determine $\phiG$ has been reported in Fig.~\ref{fig:delta_phi} {by comparing}
the density evolution of the cage order parameters with the theoretical predictions in $d\to\io$. 
%Figure~\ref{fig:delta_phi} shows that $\Delta_{AB}$ and $\Delta$ coincide when $\f < \phiG$ but separate around $\phiG$. 
However, while the effect is qualitatively {(and visually) clear, obtaining a quantitative measure of $\phiG$ from this approach is somewhat ill-defined.}
{For $\f > \phiG$ one also should treat $\Delta$ results with caution.} As discussed above, because of the finite-size and out-of-equilibrium nature of the system, the value of $\Delta$ drifts with time as different sub-basins are explored.
%If we define $\Delta$ as the limit $t\to\io$ \emph{after} taking the thermodynamic limit $N \to \infty$, i.e., prevent any activations between metabasins/sub-basins other than through hopping, then we should observe $\Delta=\Delta_1$ in the 1RSB phase and $\Delta_{\rm EA}$ in the fullRSB phase. If, however, we take the limit $t \to \infty$ \emph{before} sending $N \to \infty$, i.e., allow activations between sub-basins but not between metabasins other than through hopping, then $\Delta$ should correspond to the restricted equilibrium values $\Delta_{\rm SF} \equiv \int \Delta P_{\rm SF}(\Delta) d\Delta$ in both phases (see Fig.~\ref{fig:fractal}). Because in the MK model free energy barriers scale with $N$, we do not expect activation to be insignificant in the Gardner phase.
{Because in practice} we evaluate $\Delta$ at a relatively short time $t_{\rm s}$ (Sec.~\ref{sec:static}), {we likely} obtain a reasonable estimate of the size of a single sub-basin, {but this procedure is also somewhat uncontrolled.}

Overall, {we find that} the most reliable way to determine $\phiG$ is the divergence of $\chi$ (procedure {\it (i)}). This {effect} is clearly
the most spectacular signature of the transition, and sample-to-sample fluctuations do not much affect {its detection} % much this procedure
(Fig.~\ref{fig:sample_dependence}). Interestingly, because $\chi_{AB}$ is almost independent of time, one can determine $\chi$ reliably
using $\DAB$ at short times, as we did in this paper.
Once $\phiG$ is determined in this way, a useful test is to check that this value is consistent
with the behavior of $\D$ and $\DAB$, as in Fig.~\ref{fig:delta_phi}.
Procedure {\it (iii)}, i.e. finding the maximum of the skewness, requires averaging
over a large number of samples which will surely be difficult in numerical simulations of realistic models of glasses as well as in experiments,
where producing equilibrium configurations is extremely difficult. Procedure {\it (i)}, i.e. the divergence of $\tau_\b$, is also difficult because
the determination of $\tau_\b$ is subject to some ambiguity, but the study of the dynamics is useful because aging effects are manifest
for $\f > \phiG$ (Fig.~\ref{fig:delta-ttw}).
Procedure {\it (iv)} is used here as a consistency check with the theory, 
rather than a method to detect $\phiG$.

\section{Conclusions}
\label{sec:conclusion}
The Gardner transition separates a stable glass {metabasin} at low densities (high temperatures) from a complex {hierarchy of marginally stable sub-basins, at high densities (low temperatures)}. Its existence, which is proven
in mean-field glass models~\cite{Ga85,GKS85}, has deep consequences on the low-temperature physics of glasses {and on jamming}~\cite{CKPUZ14,CCPZ15}. It is therefore
extremely interesting to check whether such a transition exists in realistic models of glasses and in experiments.

In this work we have investigated and compared several numerical procedures for detecting the Gardner
transition in the MK model, which belongs to the universality class of mean-field {spin glasses} while {remaining fairly close} to realistic glass formers~\cite{MK11}.
We have presented three independent
approaches for locating the transition, 
all of which show that the transition 
exists and is found in a region that is roughly consistent with theoretical predictions. We have discussed the advantages and drawbacks of each of these strategies
and the {importance} of finite-size effects.

This work paves the way for studying the Gardner transition in more
realistic numerical models of glasses, where the very existence of the Gardner transition is
debated~\cite{UB14}. 
Our approach is also suitable to be
reproduced in experiments.
SF, for instance, corresponds to a straightforward annealing, and some of the observables
should be readily available through standard microscopy or scattering techniques.

One key hurdle to generalizing our methodology to other systems is the 
need to
equilibrate, without planting (and thus through slow annealing), a glass state well above the (avoided) dynamic glass transition.
To follow adiabatically a glass state, one should be able to prepare
initial configurations such that the $\a$-relaxation time $\t_\a$ is very large ($\t_\a \gg \t_\b$), 
so that the SF experiment can be performed on time scale $\t_\b \ll \t \ll \t_\a$, as discussed in Sec.~\ref{sec:timescales}.
For numerical simulations this requirement can be particularly computationally onerous, but it may be more easily achievable in experimental systems, where longer timescales can typically be reached.
In particular, it would be very interesting to investigate the existence of the Gardner transition in ultrastable glasses that 
can be prepared through vapor deposition and have an extremely large $\t_\a$~\cite{Singh2013, Hocky2014,ultra1,ultra2}.
In experiments, the bigger challenge would be to substitute 
the cloning procedure with a (potentially very) large number of experimental replicates.

Finite-dimensional non-mean-field glass formers 
display features that are not observable in the
MK model. In particular, we expect a diverging length scale to be
associated with the Gardner transition in these systems. This length
scale is expected to capture static heterogeneity, which represents the
spatial inhomogeneity of cage sizes around and above $\phiG$.  In
principle, this kind of static heterogeneity should be different from
both the dynamic heterogeneity around the dynamic glass transition,
and the heterogeneity close to jamming, which is related to soft
relaxation modes~\cite{BW09b}. 
Understanding the relevance of marginal stability for glassy dynamics, and the relation with the
ideas of Ref.~\cite{BW09b}, would open a new window on the property of low-temperature glasses.

\acknowledgments

We wish to thank L.~Berthier, O.~Dauchot, W.~Kob, J.~P.~Bouchaud, G.~Biroli, J.~Kurchan, S.~Franz, 
T.~Rizzo, F.~Ricci-Tersenghi, and M.~Wyart for very useful discussions, and especially
P.~Urbani and H.~Yoshino for collaborating with two of us in the theoretical part of this project~\cite{Rainone2014,Rainone2666}.
P.C. acknowledges support from the Alfred P. Sloan Foundation and NSF support No. NSF DMR-1055586. B.S. acknowledges the support by MINECO (Spain) through research contract No. FIS2012-35719-C02. The research leading to these results has received funding from the European Research Council under the European Union's Seventh Framework Programme (FP7/2007-2013)/ERC grant agreement no. [247328].

\bibliography{HS,glass,Gardner}
\end{document}